\documentclass{aa}  
\usepackage{siunitx}
\usepackage{graphicx}
\usepackage{amsmath}
\mathchardef\mhyphen="2D 

\usepackage{txfonts}

\newcommand{\kms}{\mbox{$\>{\rm km\, s^{-1}}$}}

\newcommand{\kpc}{\mbox{$\>{\rm kpc}$}} 
\newcommand{\pc}{\mbox{$\>{\rm pc}$}} 
\newcommand{\Gyr}{\mbox{$\>{\rm Gyr}$}}
\newcommand{\Myr}{\mbox{$\>{\rm Myr}$}}

\newcommand{\Msun}{\>{\rm M_{\odot}}}

\newcommand\degrees{^\circ}

\newcommand{\feh}{\mbox{$\rm [Fe/H]$}}

\newcommand{\gaia}{{\it Gaia}}

\def\ie{{\it i.e.}}

\usepackage{natbib,twoopt}
\usepackage[colorlinks,breaklinks=true]{hyperref}
\hypersetup{
    citecolor={blue},
    linkcolor={blue},
    urlcolor={blue}}
\bibpunct{(}{)}{;}{a}{}{,}            
\makeatletter
  \newcommandtwoopt{\citeads}[3][][]{\href{http://adsabs.harvard.edu/abs/#3}%
    {\def\hyper@linkstart##1##2{}%
     \let\hyper@linkend\@empty\citealp[#1][#2]{#3}}}
  \newcommandtwoopt{\citepads}[3][][]{\href{http://adsabs.harvard.edu/abs/#3}%
    {\def\hyper@linkstart##1##2{}%
     \let\hyper@linkend\@empty\citep[#1][#2]{#3}}}
  \newcommandtwoopt{\citetads}[3][][]{\href{http://adsabs.harvard.edu/abs/#3}%
    {\def\hyper@linkstart##1##2{}%
     \let\hyper@linkend\@empty\citet[#1][#2]{#3}}}
  \newcommandtwoopt{\citeyearads}[3][][]%
    {\href{http://adsabs.harvard.edu/abs/#3}
    {\def\hyper@linkstart##1##2{}%
     \let\hyper@linkend\@empty\citeyear[#1][#2]{#3}}}
\makeatother

\begin{document}

   \title{The edge of the Milky Way's star-forming disc: Evidence from a 'U-shaped' stellar age profile}

      \author{Karl Fiteni
          \inst{1,2}
          \and
          Stuart Robert Anderson\inst{3}
          \and
          Victor. P. Debattista\inst{3,4}\thanks{VPD and JC supervised and contributed equally to this work.}
          \and
          Joseph Caruana\inst{2,4\star}
          \and
          Jo\~ao A. S. Amarante\inst{5,6}
          \and
          Steven Gough-Kelly\inst{3}
          \and
          Laurent Eyer\inst{7}
          \and
          Leandro {Beraldo e Silva}\inst{8,9}
          \and
          Tigran Khachaturyants\inst{10}
          \and
          Virginia Cuomo\inst{11}
          }

   \institute{Como Lake centre for AstroPhysics (CLAP), DiSAT, Università dell’Insubria, Via Valleggio 11, 22100 Como, Italy\\
   \email{karlfiteni@gmail.com}
   \and
     Institute of Space Sciences \& Astronomy, University of Malta, Msida MSD 2080, Malta
   \and
     Jeremiah Horrocks Institute, University of Lancashire, Preston PR1 2HE, UK
    \and
    Department of Physics, University of Malta, Msida MSD 2080, Malta
    \and
    Department of Astronomy, School of Physics and Astronomy,\\Shanghai Jiao Tong University, 800 Dongchuan Road, Shanghai 200240, China\\
    \email{joaoant@gmail.com}
    \and
    State Key Laboratory of Dark Matter Physics, School of Physics and Astronomy,\\Shanghai Jiao Tong University, Shanghai 200240, China 
    \and
    D\'{e}partement d'Astronomie, Universit\'{e} de Gen\`{e}ve, Chemin Pegasi 51, 1290 Versoix, Switzerland
    \and
    Steward Observatory and Department of Astronomy, University of Arizona, 933 N. Cherry Ave., Tucson, AZ 85721, USA
    \and
    Observatório Nacional, Rio de Janeiro - RJ, 20921-400, Brasil
    \and
    Department of Astronomy, Shanghai Jiao Tong University, 800 Dongchuan Road, Shanghai 200240, P.R. China
    \and
    Departamento de Astronomía, Universidad de La Serena, Av. Raúl Bitrán 1305, La Serena, Chile
    }

   \date{Received 17 November 2025 / Accepted 25 February 2026}

  \abstract 
   {We leveraged reliable age and distance estimates from LAMOST-DR3 and APOGEE-DR17+AstroNN combined with \gaia\ data to perform a detailed analysis of the stellar age distribution in the Milky Way's (MW) outer disc using giant stars. Selecting stars near the midplane ($|z|<0.3$ kpc) on near-circular orbits ($\lambda_c > 0.9$), we analysed these independent datasets that employed different age-estimation methods. Our stringent kinematic selection criteria effectively exclude halo stars, ensuring that the observed age trends reflect genuine disc properties rather than contamination from older halo populations. Our results reveal a 'U-shaped' stellar age profile, where a negative gradient in the inner disc transitions to a positive gradient in the outer disc region. We identify the minimum in the stellar age profile at $R_{\rm min}=11.28 \pm 0.58\kpc$ and $R_{\rm min}=12.15\pm 0.62\kpc$ for the APOGEE-DR17 and LAMOST-DR3 samples, respectively. Using N-body+SPH simulations, we demonstrate that $R_{\rm min}$ corresponds to the break radius in the stellar density profile ($R_{\rm br}$), marking the edge of the Galaxy's star-forming disc. This break arises from a sharp decline in the star formation rate, with the outer positive age gradient produced by the radial migration of stars born inside $R_{\rm br}$. The cessation of star formation in the outer disc might be due to several mechanisms, including the dynamical influence of the bar's outer Lindblad resonance, the onset of the Galactic warp, or thermally regulated star formation. Overall, our results support the picture that the MW has a Type II (down-bending) stellar disc with a break at $R_{\rm br} \approx 11.28-12.15 \kpc$, where the combination of star-formation cut-off and radial migration produces the observed U-shaped age profile.
   }
   

   \maketitle

\section{Introduction}

Early studies based on photographic plates found that the light profiles of disc galaxies terminate at a sharp truncation, representing the definitive `edge' of the stellar disc \citep{vanderKruit1979, Kruit+1981a, Kruit+1981c, vanderKruit1987}. However, later investigations using CCD photometry revealed that, rather than being sharply truncated, stellar discs are more accurately described by a double-exponential model, with an exponential profile up to some break, followed by an outer disc exponential with a shorter (or longer) scale length \citep{Pohlen+2002}. This was confirmed by a subsequent study based on Sloan Digital Sky Survey (SDSS) images of $90$ face-on late-type (Sb-Sdm) spiral galaxies, which found that the light profiles of stellar discs are classified as one of three distinct types: Type I, identified as a single exponential without a break; Type II, which exhibits a downward bend in the light profile; and Type III, characterised by an upward bend \citep{Pohlen+2006}. This study found that Type II discs account for roughly $60 \%$ of disc galaxies in the local Universe. Similarly, in their sample of $66$ barred early-type (S0-Sb) galaxies, \citet{Erwin+2008} found that Type~II profiles constitute around $40\%$ of their sample. While many studies based on {\it Hubble Space Telescope} ({\it HST}) imaging have since observed breaks at intermediate redshifts ($z < 1$) \citep[e.g.][]{Perez2004, Trujillo+2005, Pohlen+2006, Azzollini+2008, Bakos+2008, Borlaff+2018}, recent studies using {\it James Webb Space Telescope} ({\it JWST}) data have extended these findings, demonstrating that Type II discs comprise roughly $50\%$ of disc galaxies at $1 < z < 3$ \citep{Xu+2024}. Observational evidence also suggests that break radii of discs increase over time, as a result of the inside-out growth of discs \citep{Trujillo+2005, Azzollini+2008}. 

While breaks among disc galaxies are common, whether the Milky Way (MW) has such a broken profile remains an open question. This is partially due to our position within the Galactic disc, which makes the stellar density profile difficult to determine directly due to dust extinction. Nonetheless, early studies based largely on star counts in the anti-centre direction have sought to constrain the MW's break radius, $R_{\rm T}$, with measurements, finding $R_{\rm T} \approx 12 - 15 \kpc$ \citep{Robin+1992, Ruphy+1996, Freudenreich+1998}. 
Using red clump (RC) giants identified in the near-infrared surveys UKIDSS-GPS and VISTA Variables in the Via Lactea (VVV), \citet{Minniti+2011} reported $R_{\rm T} = 13.9\pm 0.5 \kpc$ along various lines of sight in the Galaxy. While the authors interpreted the apparent cut-off in stellar density as a sharp truncation, they also acknowledged that a Type~II break in the distribution would also be consistent with the observed data. Similarly, \citet{Sale+2010} used a sample of $40,000$ young ($\approx 100\Myr$) A-type stars in the Galactic longitude range $160\degrees<\ell<200\degrees$ and latitude range $-1\degrees<b<1\degrees$ and obtained a value of $R_{T} = 13.0\pm 0.6 \kpc$. In contrast to the earlier studies, which favoured a truncated disc, \citet{Sale+2010} concluded that their data are more accurately represented by a Type~II break in the stellar density profile. A broken stellar density profile is also supported by studies based on mono-abundance populations (MAPs) from the APOGEE survey, which find an outermost break in the stellar density profile in the $R_{\rm br} = 10-13\kpc$ range in the low-$\alpha$, low-\feh\ disc \citep{Bovy+2016, Mackereth+2017, Lian+2022}.

Type II breaks are widely understood to be associated with star-formation thresholds \citep{Kennicutt+1989, Schaye+2004, Elmegren+2006, Martin+2012}. Numerical simulations of isolated galactic discs support the notion that Type II profiles originate from a combination of a cut-off in star-formation, coupled with radial migration driven by bars or transient spiral arms \citep{Debattista+2006, Roskar+2008b, Minchev*2011, Minchev*2012, Roskar+2012}. 
This mechanism predicts that the outer disc region beyond the break is populated by stars that migrated from the inner disc. Due to the stochastic nature of radial migration, where stars can move both outward and inward, it takes an increasingly long time for stars to reach greater radii. Specifically, the distance a star traverses scales with time as $t^{0.5}$ \citep{Sellwood+2002}. Consequently, this phenomenon leads to an increasing mean stellar age beyond the break radius. The transition from a negative age gradient inside the break radius --caused by inside-out disc growth-- to a positive stellar age gradient beyond the break creates a characteristic 'U-shaped' mean stellar age profile \citep{Roskar+2008a}. 

Consistent with this prediction, observational studies have demonstrated that Type II disc galaxies exhibit U-shaped colour and age profiles \citep{Bakos+2008, Radburn+2012, Yoachim+2012}. Using photometry for 85 late-type spiral galaxies from the SDSS, \citet{Bakos+2008} found that galaxies with Type~II breaks exhibited U-shaped $(g'-r')$ colour profiles, with the minimum coinciding with the break in the light profiles. \citet{Yoachim+2012} used spatially resolved spectroscopic data of 12 nearby disc galaxies to derive the stellar age profiles in their outer discs. They observed that in three instances, a downward break corresponds to an increase in the stellar age in the outer disc, while the inner disc remains dominated by ongoing star formation. Conversely, in three other cases, they found no increase in age beyond the break, indicating that both the inner and outer discs are characterised by active star formation. Using photometric data of resolved stars from Hubble Space Telescope (HST) observations, \citet{Radburn+2012} analysed distinct stellar populations covering a wide range of ages within the spiral galaxy NGC 7793. They found that beyond the break radius, the steepness of the radial surface brightness profile increases for younger stellar populations, which is consistent with the effects of radial migration. 

Whether an age upturn in the stellar age profile is present in the MW's outer disc remains uncertain. The first hints of an upturn came from \citet{Wu*2019}, which presented a catalogue of stellar age and mass estimates for over $640$k RGB stars from LAMOST-DR4 and found that the median stellar age profile for stars in the midplane ($|z|<0.5\kpc$) starts to show a trend towards older stars beyond $12\kpc$ (their Fig. 16). Using data from APOGEE-DR16, \citet{Lian+2022} independently found that beyond $13 \kpc$, stars show a trend towards older ages (see their Figure 1), which they attribute to radial migration. 

The ability of radial migration to reconfigure the stellar populations within a disc has been demonstrated in simulations, with studies finding that migrated stars constitute a large fraction (above $50\%$) of the stellar population in the outer disc \citep[e.g.][]{Roskar+2008a,Martinez+2009,Roskar+2012, Vera-ciro+2014, Loebman+2016}. Strong evidence for radial migration in the MW has also been obtained, largely through studies of the age-metallicity relation (AMR) \citep{Feltzing+2001,Nordstrom+2004,Xiang+2017,Mackereth+2017,Silva+2018,Hasselquist+2019, Haywood+2024, Zhang+2025, Ratcliffe+2025}, the stellar metallicity distribution function (MDF) \citep{Kordopatis+2015,Halle+2015}, and stellar kinematics \citep{BeS+2021, Dillamore+2024}. Using APOGEE-DR12 data, \citet{Hayden+2015} found that the MDF is negatively skewed at small radii (3-7\kpc) and positively skewed beyond the solar radius -- a signature that they attributed to radial migration. \citet{Loebman+2016} confirmed this using N-body+SPH simulations, which closely match the observed MDF skewness as a function of radius. They concluded that the migrated fraction increases with radius, with about half of stars at the solar cylinder having migrated from elsewhere. \citet{Sharma+2021} independently found the same skewness in the MDF in a later APOGEE-DR14 release. Using data from RAVE-DR4, \citet{Kordopatis+2015} found that roughly half of the stars in the Solar neighbourhood with super-solar metallicity are on circular orbits and suggested that they must have migrated from the inner disc. Analysing a sample of old stars in the solar neighbourhood, \cite{BeS+2021} found evidence of radial migration affecting both the [$\alpha$/Fe]-poor and [$\alpha$/Fe]-rich discs, with $\approx 50\%$ of the [$\alpha$/Fe]-poor stars in the sample migrating from the inner disc, and $\approx 30\%$ of the [$\alpha$/Fe]-rich stars doing the same. Radial migration has also been linked to changes in the observed $\feh-R$ gradient in the outer MW disc. 

\begin{figure*}
    \centering
    \includegraphics[width=.9\linewidth]{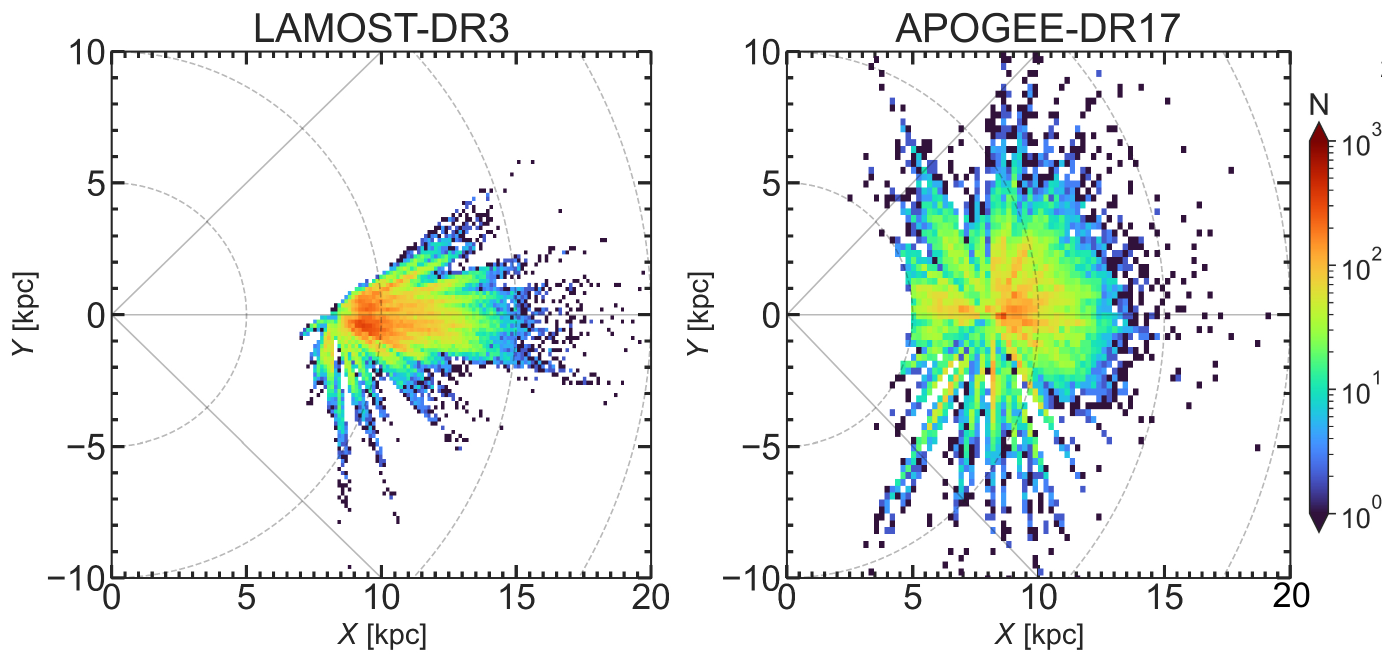}
    \caption{2D histograms representing the spatial coverage in galactocentric coordinates of the LAMOST-DR3 (left) and APOGEE-DR17 (right) samples. The two samples assume slightly different solar positions: $(X, Y, Z) = (8.2,0,0.015)$ kpc and $(8.125,0,0.021)$ kpc for LAMOST-DR3 and APOGEE-DR17, respectively.}
    \label{fig:datasets_xy}
\end{figure*}

In this study, we used stellar age measurements from LAMOST-DR3 and APOGEE-DR17, coupled with \gaia\ astrometry, to reveal a positive stellar age gradient in the MW's outer geometric thin disc. Using these samples, we constrain the location of the break within the MW stellar disc. Throughout this paper, when discussing radial migration, we refer exclusively to churning (or cold torquing), defined as a change in the guiding radius, $R_g$ (or $L_z$) of a star's orbit, with minimal changes to its radial action, $J_R$ \citep{Sellwood+2002}. This is in contrast to orbital heating (or blurring), which is characterised by an increase in the orbital eccentricity (or $J_R$); see, for example, \citet{Schonrich+2009}.

This paper is structured as follows. Section \ref{sec:datasets} provides an overview of the observational data employed to analyse the stellar age distribution in the MW disc. In Section \ref{sec:profile_mw}, we present the stellar age profiles derived from these observational datasets. Section \ref{sec:model_profiles} focuses on comparing the observational results with N-body+SPH models. We explore the implications of our findings in Section \ref{sec:discussion}. Finally, we summarise our results in Section \ref{sec:summary}.

\section{Observational samples}\label{sec:datasets}

\begin{figure*}
        \includegraphics[width=\linewidth]{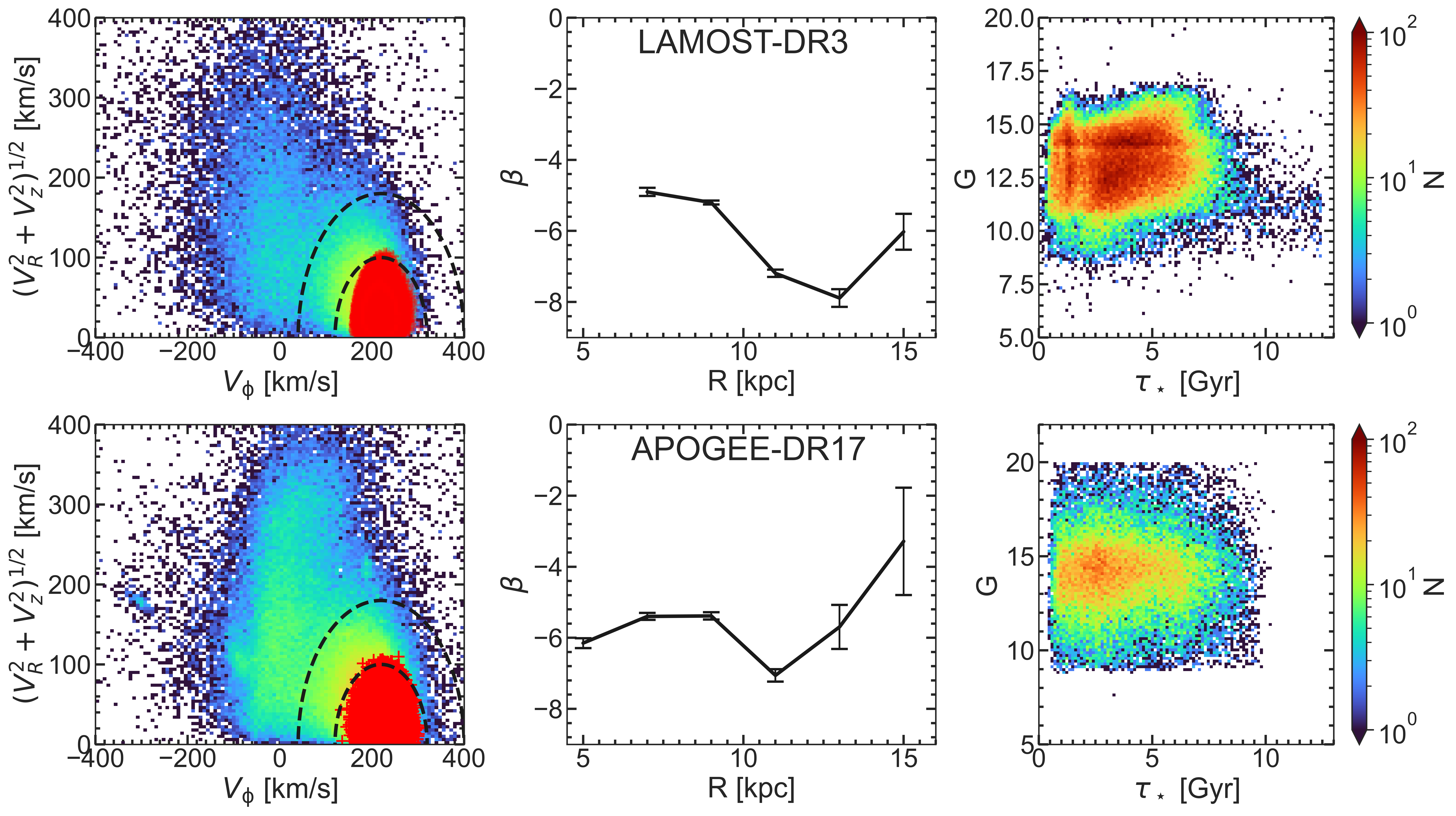}
    \caption{Left: Toomre diagrams showing the kinematics of the uncleaned LAMOST-DR3 (top row) and APOGEE-DR17 (bottom row) catalogues, (blue/green distribution) with the stars satisfying $\lambda_{\rm c} > 0.9$ and $|z|<0.3\kpc$ over-plotted with red markers. The inner and outer dashed curves represent the boundaries between the kinematically defined thin/thick discs and halo. Middle: Orbital anisotropy, $\beta$, of the cleaned sub-sample with errors determined via a bootstrapping method. The negative value of $\beta$ indicates that the population is tangentially biased. Right: Distribution of apparent \gaia\ $G$-band magnitude versus stellar age, $\tau_{\star}$, for the cleaned sub-samples. }
    \label{fig:lamost_chr}
\end{figure*}

\begin{figure}
    \hspace{.3cm}
    \includegraphics[width=\linewidth]{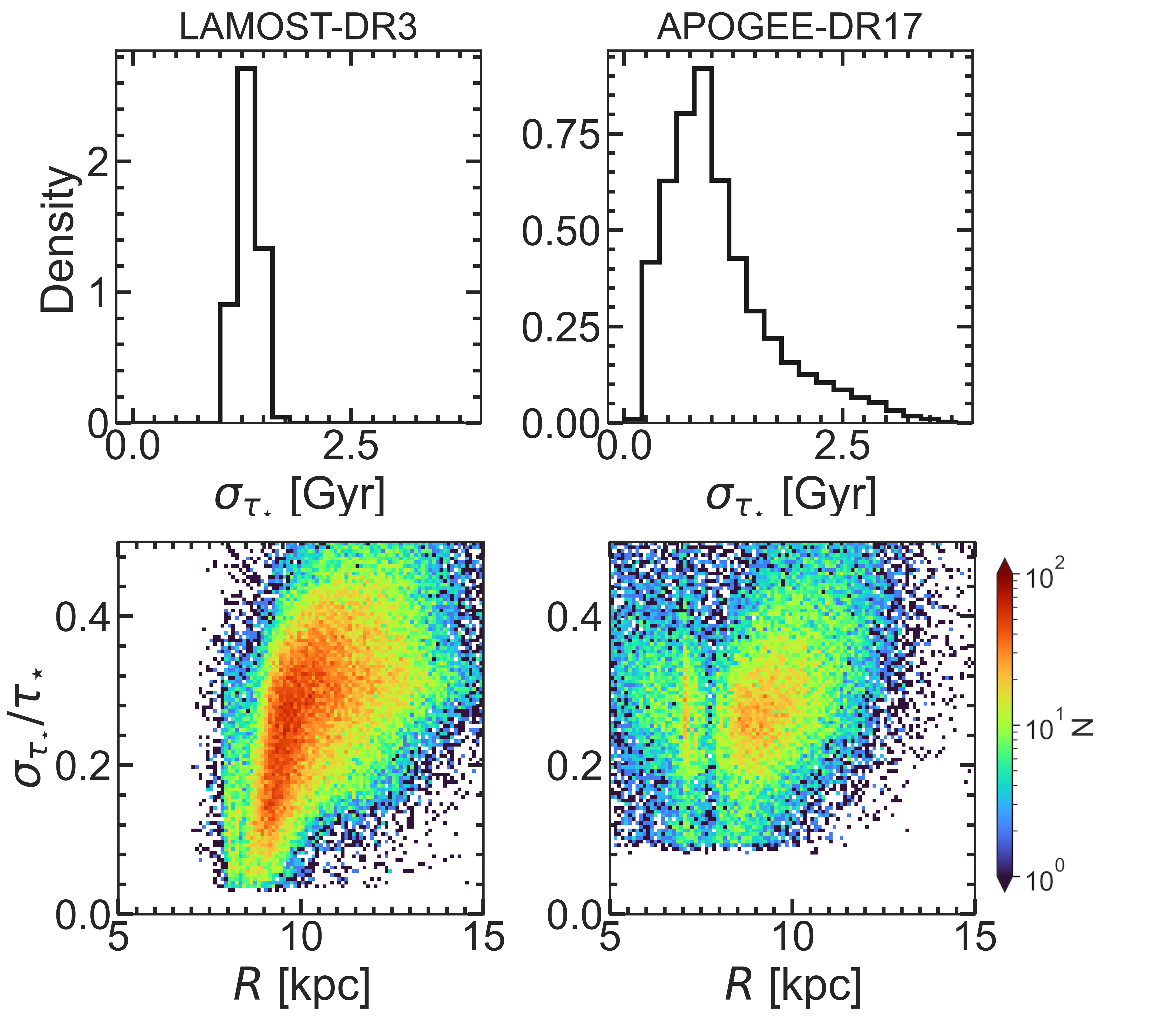}
    \caption{Top row shows the age error distribution for our cleaned LAMOST-DR3 (left) and APOGEE-DR17 (right) samples.
     The bottom row shows the 2D histograms of the relative age error versus $R$ for the two samples.}
    \label{fig:age_error_dist}
\end{figure}

We used two independent samples with full 6D phase-space coordinates and stellar ages, $\tau_{ \star}$, estimated via independent methods. Although several catalogues providing stellar ages are available, our selection of these datasets is primarily driven by their spatial coverage, particularly towards the Galactic anti-centre. To reduce contamination from halo stars and the influence on the results from orbital heating, we restricted our analysis to stars that are close to the Galactic mid-plane ($|z| < 0.3\kpc$) and are on near-circular orbits. For the orbital selection, we impose a criterion on the circularity parameter, $\lambda_{c} = L_{z}(E)/L_c(E) > 0.9$, using the python library {\sc agama} \citep{Vasiliev2019} to integrate the orbits of stars in the Galactic potential of \citet{Mcmillan2017}. Here, $L_z(E)$ is the angular momentum of a star with energy, $E$, and $L_c$ is the angular momentum of a star on a circular orbit having the same energy. Thus, $\lambda_c = 1$ corresponds to a star on a circular orbit in the Galactic plane.

\subsection{LAMOST-DR3}\label{sec:sandas}

Our first sample is based on 6D \gaia-DR2 coordinates and age estimates from \citet{Sanders+2018}. They used a Bayesian framework for isochrone fitting in order to characterise the probability density functions for distance, mass, and age using photometric, spectroscopic, and astrometric data. \citet{Wang+2023} point out that deriving stellar ages via isochrone fitting is challenging due to the possible mixing of stars in different evolutionary stages in the Hertzsprung--Russell diagram. However, \citet{Wang+2023} also show broad agreement between their ages --derived via a neural network method-- and those of \citet{Sanders+2018}.

While the catalogue of \citet{Sanders+2018} combines data from multiple spectroscopic surveys, we only used stars with spectroscopic data from LAMOST-DR3 \citep{Cui+2012, Zhao+2012}, which is a low-resolution ($R \sim 1800$) optical ($3650-9000\,\si{\angstrom}$) survey dedicated to exploring the disc and halo. We used this survey as it provides the best coverage of the outer disc region. \citet{Sanders+2018} used the reported effective temperature, $T_{\rm eff}$; surface gravity, $\log g;$ and [Fe/H] from LAMOST-DR3, complemented by a combination of 2MASS or Pan-STARRS photometry. We used the sub-sample of giant stars, which the authors defined as stars having $\log g < 3$ dex and $\log_{10}(T_{\rm eff}/K)<3.73$. Because the masses of giants are closely correlated with their ages, \citet{Sanders+2018} incorporated [C/N] abundances into their Bayesian pipeline to accurately determine mass and subsequently provide precise age estimates for the giants. We applied quality cuts to the giants by requiring that flag {\fontfamily{qcr}\selectfont Best\,$=1$} and  {\fontfamily{qcr}\selectfont ruwe\,$<1.4$}. The {\fontfamily{qcr}\selectfont Best\,$=1$} flag ensures that the isochrone pipeline --which estimates ages-- was successful and that no issues with the input spectroscopy, photometry, astrometry, or mass arose. This also removes a number of red dwarfs with very young ages ($\tau_{\star}< 100$ Myr). The cut on {\fontfamily{qcr}\selectfont ruwe\,$<1.4$}, which we acquired by cross-matching the sample with the \gaia-DR2 source catalogue, ensures that the target star has a reliable astrometric solution in the \gaia\ pipeline and removes potential binaries. We made another cut on the relative parallax error ($\sigma_{\varpi} / \varpi < 0.2$) and relative age error, $\sigma_{\tau_{\star}}/\tau_{\star} <0.5$. As suggested by \citet{Sanders+2018}, we made a further cut on the age uncertainty ($\sigma_{\log{\tau_{\star}}}>0.015$ dex). This left us with a total of $66,615$ kinematically cool giants from LAMOST-DR3. The average age uncertainty, $\left<\sigma_{\tau_{\star}}\right>$, for this sample is $1.32\Gyr$.

\begin{figure}
        \includegraphics[width=\linewidth]{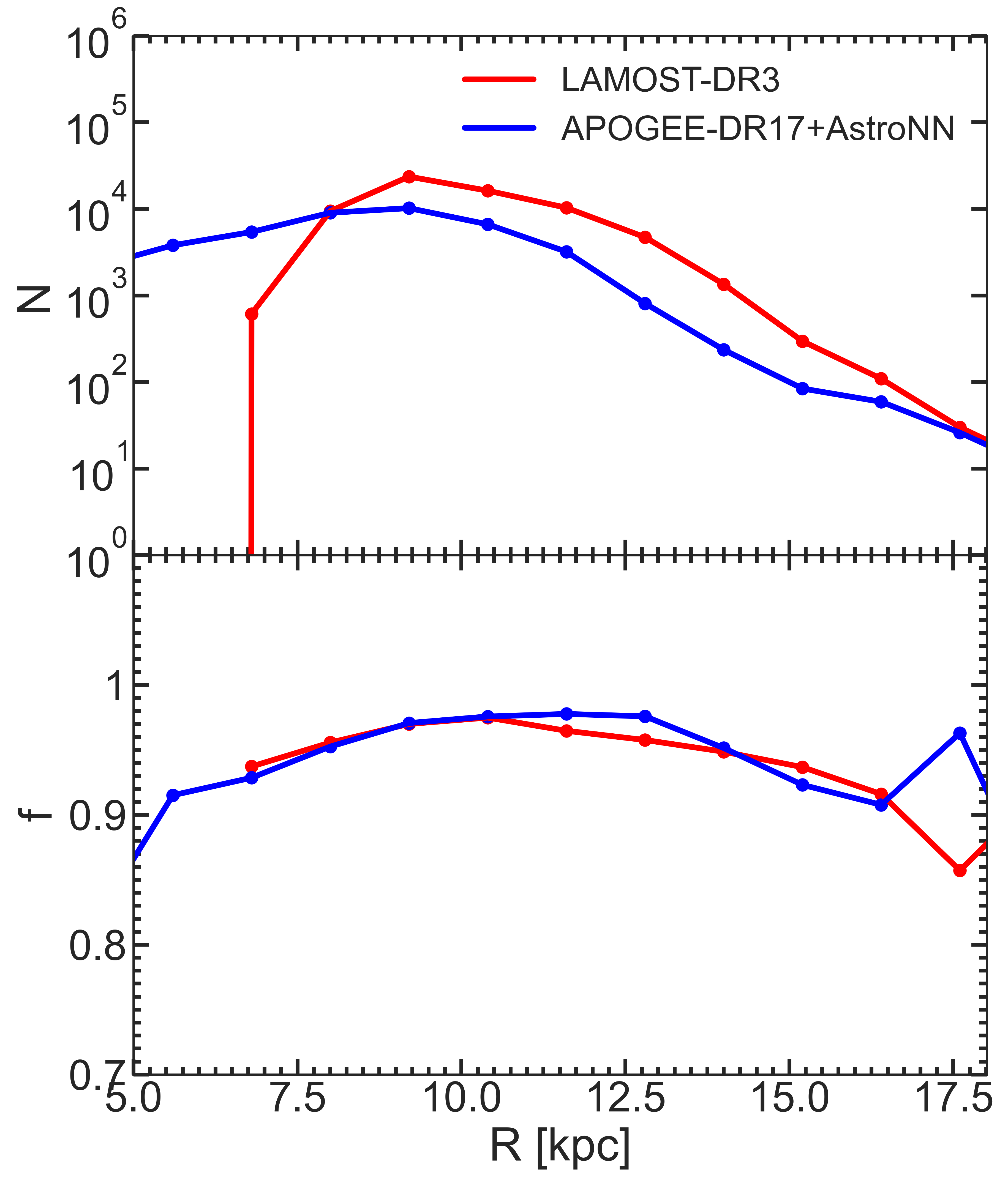}
    \caption{Top panel: Number distribution of selected stars as a function of $R$ for the cleaned LAMOST-DR3 (red) and APOGEE-DR17 (blue) samples. Bottom panel: Fraction of stars retained after applying the additional cut in circularity $(\lambda_{c}>0.9)$ on stars in the mid-plane $(|z|<0.3\kpc)$.}
    \label{fig:star_counts}
\end{figure}

\subsection{APOGEE-DR17+AstroNN}\label{sec:apogee}

The APOGEE sample is built from the cross-match between Apache Point Observatory Galactic Evolution Experiment-2 (APOGEE-2) observations contained in the Sloan Digital Sky Survey-IV (SDSS-IV) Data Release 17 (DR17) \citep{Blanton+2017,Majewski+2017,Abdurro+2022} and \gaia-EDR3 data \citep{gaia1, gaiaEDR3}. We first selected stars with the relative parallax error $\sigma_{\varpi} / \varpi < 0.2$. We used the abundances and stellar parameters derived from the APOGEE Stellar Parameters and Abundances Pipeline \citep[ASPCAP][]{Nidever+2015}. As in previous works \citep[e.g.][]{Lian+2021,Lian+2022, leung+2022}, we selected giant stars with signal-to-noise ratios of $>70$, $T_{\mathrm{eff}} > 3500K$, $0.8 < \log(g) < 3.5$, and $\sigma_{\log(g)} < 0.15$. We applied quality flags to ensure that the target belongs to the main survey and to remove duplicate stars (EXTRATARG~==~0). Additionally, we eliminated possible stellar clusters based on the APOGEE1\_TARGET1 and APOGEE2\_TARGET1 flags. We also applied the bitmasks\footnote{\url{https://www.sdss.org/dr17/algorithms/bitmasks/}} 4,~9,~16,~17~==~0 in STARFLAG, and 19,~23~==~0 in ASPCAP, thus removing stars with warnings in their parameter measurements. As we are interested in exploring the outer disc region, we also removed all stars with $R < 5\kpc$.

The ages and distances of our cleaned sample are taken from the {\tt astroNN} \citep{leung+2019,mackereth+2019} value added catalogue\footnote{\url{https://data.sdss.org/datamodel/files/APOGEE_ASTRONN/apogee_astronn.html}}, which provides stellar ages and distances derived through a neural network framework described in detail in \citet{leung+2019}, and retrained for APOGEE-DR17 \citep{leung+2022}. Finally, we selected stars with a relative age error of $\sigma_{\tau_{\star}}/\tau_{\star} <0.5$. This left us with a sample of $35, 507$ stars with an average age uncertainty of $\left<\sigma_{\tau_{\star}}\right> = 1.12$ Gyr.

We also verified that for both the LAMOST-DR3 and the APOGEE-DR17 samples, our results were qualitatively consistent when applying a more stringent cut for the relative age error of $\sigma_{\tau_{\star}}/\tau_{\star} <0.3$ (see Fig.~\ref{fig:mw_fit_30}). Although this approach yielded more reliable age estimates, we chose to prioritise spatial coverage in the disc's anti-centre by adopting a less restrictive cut for the age error.

\begin{figure*}
    \includegraphics[width=\linewidth]{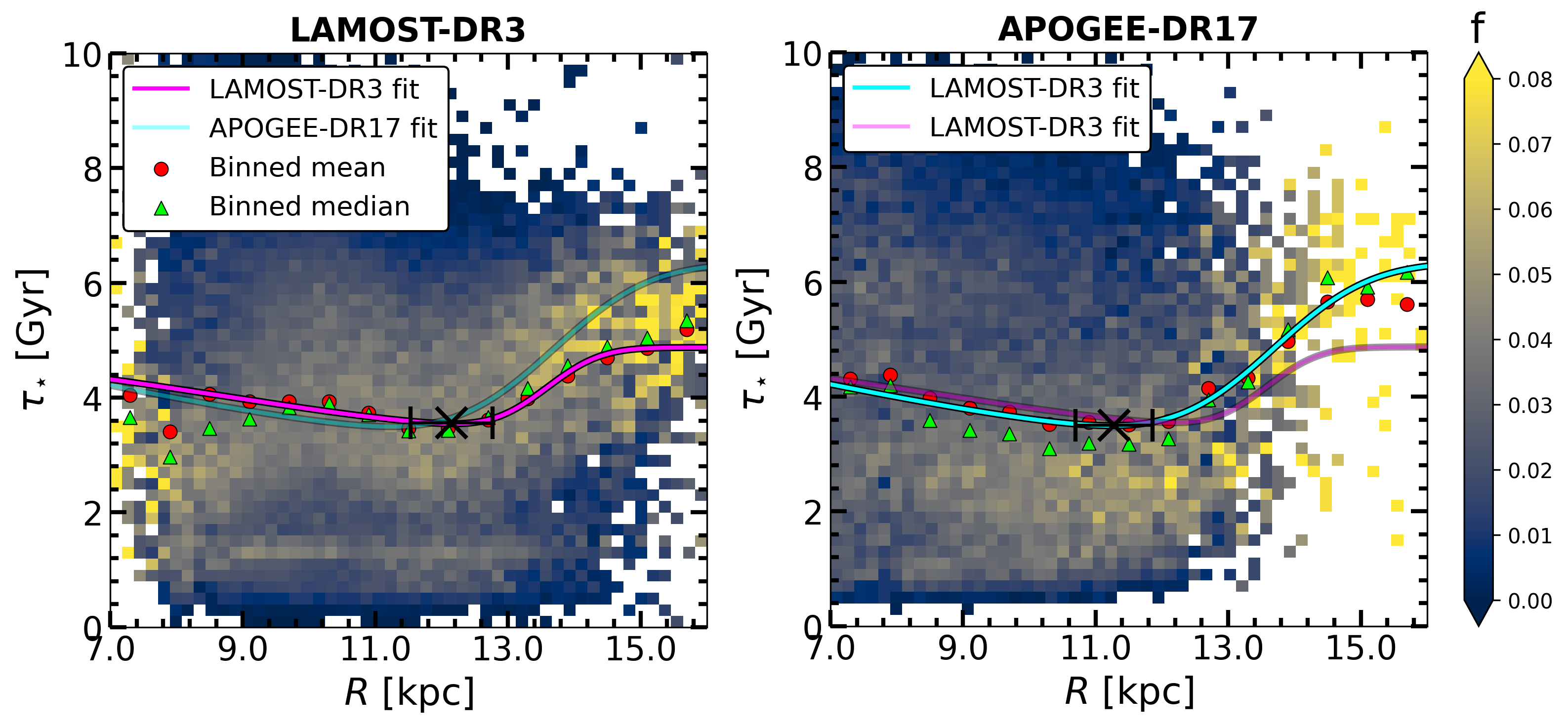}
    \caption{Column-normalised 2D histograms of the stellar age distribution, $\tau_{\star}(R)$, for the LAMOST-DR3 (left) and APOGEE-DR17 (right) samples. The red and green markers show the mean and median profiles, respectively. The solid curves show the age profile resulting from the maximum-likelihood fitting (this fits to all individual sources without binning -- see text for details). In each panel, we also cross-plot the ML fit of the dataset in the opposite panel for comparison. The black marker reflects the location of the maximum-likelihood age minimum and associated uncertainty.}
    \label{fig:obs_fit_ML}
\end{figure*}

\subsection{Observational sample properties}\label{sec:sample_properties}

In the left and right panels of Fig.~\ref{fig:datasets_xy}, we present the observational footprints for the cleaned LAMOST-DR3 and APOGEE-DR17 datasets, respectively. The majority of stars in the LAMOST-DR3 dataset are concentrated in the anti-centre direction, covering a (galactocentric) radial distance of ${7<R\,/\kpc<20}$. On the other hand, APOGEE-DR17 encompasses a broader azimuthal coverage, with fewer stars beyond $R = 15 \kpc$ compared to LAMOST-DR3.

In Fig.~\ref{fig:lamost_chr}, we summarise properties of the LAMOST-DR3 (top row) and APOGEE-DR17 (bottom row) samples. The left column shows Toomre diagrams, where the blue/green density distribution reflects the entire population of the respective samples, and the red markers represent the cleaned (i.e.\ $|z| < 0.3 \kpc$ and $\lambda_{\rm c} > 0.9$) samples we used for the rest of the analysis. Here, $V_R$, $V_{\phi}$, and $V_z$ are the cylindrical galactocentric velocities given by the catalogues. In order to recover the galactocentric parameters, both samples used the peculiar solar velocity from \citet{Schonrich+2010}\footnote{$(U_{\odot},V_{\odot},W_{\odot})=(11.1,12.24,7.25)\kms$}. However, while the LAMOST-DR3 sample assumed a solar position of $R_0 = 8.2\kpc$ and $z_0=15\pc$ \citep{Bland+2016}, the APOGEE-DR17 sample assumed $R_0 = 8.125 \kpc$ \citep{Gravity+2018} and $z_0 = 21\pc$ \citep{Bennett+2019}. 

Halo stars are generally characterised by older stellar ages. Therefore, verifying that our selection criteria effectively exclude them is crucial to ensure that observed variations in stellar age trends in the outer disc are not influenced by their contamination. To demonstrate this, we used Toomre diagrams (left column in Fig.~\ref{fig:lamost_chr}), which were used to kinematically distinguish between the major stellar components of the MW (i.e.\ thin/thick discs and halo) \citep[e.g.][]{Venn+2004,Altmann+2004,Qu+2011,Yepeng+2019}, with the cuts based on the velocity relative to the Local Standard of Rest (LSR). We represent the boundaries between the kinematically defined components with dashed black curves. Stars with $|V-V_{\odot}| <100\kms$ and $100\kms <| V-V_{\odot} |< 180\kms$ are associated with the thin and the thick discs, respectively \citep[e.g.][]{Venn+2004, Bensby+2014, Limberg+2021}. Beyond $| V-V_{\odot} |> 180\kms$, stars are associated with the halo, which comprises a number of substructures, including heated in situ and accreted populations, reflecting the MW's turbulent early history \citep[e.g.][]{Belokurov+2018,Helmi+2018,Haywood+2018,DiMatteo+2019,Belokurov+2020,Amarante+2020}. In both clean samples, the sources are centred around $V_{\rm \phi} \approx 220\kms$ and lie within $|V-V_{\odot}| <100\kms$, which is the region kinematically defined as the thin disc. There is also a negligible number of stars that lie in the thick disc region. However, the cuts we applied (i.e.\  $|z| < 0.3\kpc$ and $\lambda_{c} = L_{z}/L_c(E) > 0.9$) successfully removed any stars belonging to the halo.

The middle column of Fig.~\ref{fig:lamost_chr} shows the orbital anisotropy, $\beta$, as a function of galactocentric radius, $R$, for the cleaned samples. Here, $\beta = 1-\sigma_{\phi}^2/\sigma_{R}^2$, where a value of $0< \beta \leq 1$ indicates a stellar population that is radially biased, typical of a halo-like population. On the other hand, $- \infty \leq \beta < 0$ indicates a tangentially biased population and is characteristic of nearly circular disc orbits. The uncertainty for the anisotropy was calculated using a bootstrap method, which involved resampling the stars with replacement within each radial bin for $1000$ iterations and adopting the standard deviation as the uncertainty. The negative values for $\beta$ at all radii in our cleaned samples, combined with the results from the Toomre diagram, confirm that stars in the outer disc are not an older population of halo stars, which would bias our later analysis. The right column of Fig.~\ref{fig:lamost_chr} shows the \gaia\ $G$-band magnitude versus stellar age, $\tau_{\star}$, for the cleaned samples, showing that the samples are not biased by older and brighter sources. 

The top row of Fig.~\ref{fig:age_error_dist} shows the distributions of the uncertainties in the stellar age, $\sigma_{\tau_\star}$, for the two samples. The bottom row plots the 2D histogram of the relative age errors, $\sigma_{\tau_{\star}}/\tau_{\star}$ versus $R$, coloured by the number of stars in our clean samples of LAMOST-DR3 (left) and APOGEE-DR17 (right). As expected, the relative error increases with $R$.

The top panel of Fig.~\ref{fig:star_counts} shows the star counts, $N$, versus $R$ for stars satisfying $|z| < 0.3\kpc$ and $\lambda_c>0.9$ in the LAMOST-DR3 (red) and APOGEE-DR17 (blue) surveys. The bottom panel shows the fraction of disc stars (at $|z| < 0.3\kpc$) that are retained when applying the additional cut in circularity ($\lambda_c>0.9$). Throughout the disc region, the fraction remains consistently high ($f\gtrsim0.90$), indicating that a substantial portion of stars in the outer MW disc are on cool orbits, which is consistent with previous studies of the MW \citep{Frankel+2020,Lian+2022,Hamilton+2024}. 

\begin{figure*}

    \includegraphics[width=\linewidth]{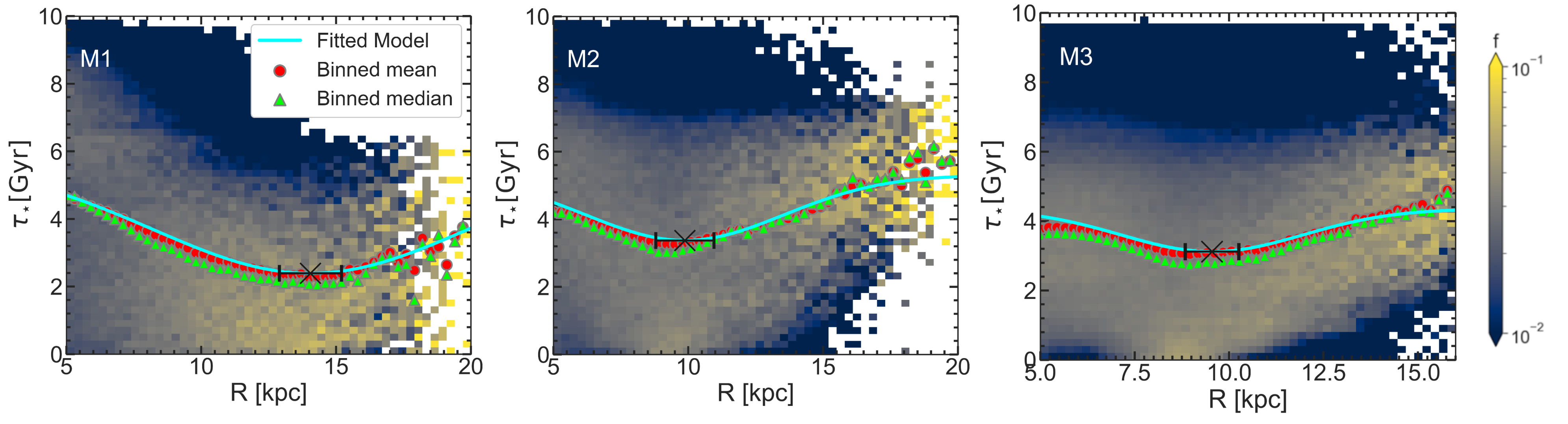}
    \caption{Column-normalised stellar age distribution of the models. The red and green marker points show the mean and median stellar age profiles, respectively. The solid cyan line shows the final age profile, $\tau_{\star}$, resulting from the ML fitting. The black point shows the position and associated uncertainty of $R_{\rm min}$ as determined by $\tau_{\star}^{\prime}=0$.}
    \label{fig:age_fits_all_models}
\end{figure*}

\section{The MW stellar age profile}\label{sec:profile_mw}

The radial stellar age profiles derived from the LAMOST-DR3 and APOGEE-DR17 samples are shown in Fig.~\ref{fig:obs_fit_ML} (magenta and cyan curves in the left and right panels, respectively). Here, we give a brief outline of the method used to generate the profiles, with full details provided in Appendix~\ref{sec:age_profile_modelling}. Our approach employs a maximum-likelihood (ML) method, which fits a function directly to the individual sources, rather than using bins in radius and age, thereby avoiding any imposed biases associated with binning choices. The solid curves in  Fig.~\ref{fig:obs_fit_ML} show the fitted age profiles, $\tau_{\star}(R)$. For comparison, each panel also displays the fit from the opposite dataset. In order to reliably model the stellar age profiles, we evaluated several functional forms. Each function presented advantages and limitations in fitting the data. For our analysis, we prioritised functions that accurately model the region surrounding the age minimum. For the observational data, the fits follow an inverted skewed Gaussian of the following form:
\begin{equation}
    \tau_{\star}(R)=-a\exp[-c(R-b)^2][1+\mathrm{erf}(d(R-b))] + \epsilon ,
    \label{eqn:skewed-Gaussian2}
\end{equation}
with $\tau_{\star}$ in gigayears and $R$ in kpc, and $a, b, c, d, \epsilon$ are free parameters\footnote{When $d=0$, the function reduces to a symmetric Gaussian.}. In order to obtain a reasonable initial guess for the fit parameters, we first fitted Equation~\ref{eqn:skewed-Gaussian2} to the mean age profile binned in $R$ (red circles). For reference only, we also plot the column-normalised 2D histogram in the $(\tau_{\star},R)$ plane. We identify the minimum point in the age profile, $R_{\rm min}$, as the point satisfying the condition $\tau^{\prime}_{\star}(R)=d\tau_{\star}(R)/dR=0$. 

To determine the uncertainty in $R_{\rm min}$, we used a curvature-based approach that accounts for both the data scatter and the sharpness of the fitted age profile. The uncertainty is given by $\sigma_{R} = \sqrt{2\sigma_{\rm age}/|f''(R_{\rm min})|}$, where $\sigma_{\rm age}$ is the standard deviation of residuals between the individual stellar ages and the ML-fitted profile, and $f''(R_{\rm min})$ is the second derivative of the fitted function (Eqn.~\ref{eqn:skewed-Gaussian2}) evaluated at $R_{\rm min}$. This method directly relates the positional uncertainty to the curvature of the age profile: sharper minima yield smaller uncertainties. We also tested bootstrap resampling, but found it yielded unrealistically small uncertainties; we therefore adopted this curvature-based method as it properly reflects the shape of the age profile around $R_{\rm min}$.

The two independent samples both show a negative age gradient in the inner disc $(R \leq 10 \kpc)$, which is consistent with the inside-out growth scenario for the MW thin disc \citep[e.g.][]{Matteucci+1989, bird+2013, Frankel+2019, Katz+2021, Ratcliffe+2025}. In contrast, a positive age gradient is observed in the outer disc of both samples $(R \geq 12.0 \kpc)$. The maximum-likelihood profiles for the LAMOST-DR3 dataset (magenta left) yields an age upturn at $R_{\rm min}=12.15\pm0.62\kpc$. Meanwhile, the APOGEE-DR17 dataset (cyan right) yields an age upturn at $R_{\rm min} = 11.28\pm 0.58\kpc$. These results (summarised in Table~\ref{table:MW_age_prof_params}) are consistent between the two datasets within the uncertainties.

\begin{table}
\caption{Cylindrical radius of the age minimum $(R_{\rm min})$ for the observational datasets and simulations.}
\label{table:MW_age_prof_params}
\centering
\renewcommand{\arraystretch}{1.2}
\begin{tabular}{lc}
\hline\hline
Dataset / Model & $R_{\rm min}$ [kpc] \\
\hline
LAMOST-DR3  & $12.15\pm0.62$ \\
APOGEE-DR17 & $11.28\pm0.58$ \\
M1          & $14.04\pm1.15$ \\
M2          & $9.87\pm1.08$  \\
M3          & $9.54\pm0.71$  \\
\hline
\end{tabular}
\end{table}

\section{Comparison with simulations}\label{sec:model_profiles}

We employed three N-body+SPH models to investigate the radial age profiles of stellar populations. The models were evolved for $10\Gyr$ using {\sc gasoline} \citep{Wadsleyetal2004, Wadsley2017}. Models M1 and M2 are designed to simulate conditions following an early merger. The merger models are part of the {\sc gastro} simulation suite \citep{amarante+2022, Amarante+2025}, which is specifically designed to explore the effects of a single merger event, including \gaia-Sausage-Enceladus (GSE) on the MW \citep[e.g.][]{Belokurov+2018, Helmi+2018}. Each merger simulation features variations in the initial conditions of the satellite and different sub-grid physics (see Section~\ref{sec:model_gastro}). For all the models, the merger is completed in the first $~3\Gyr$ of evolution. Model M3 simulates an isolated disc (see Section~\ref{sec:model_iso} for details). In order to be consistent with the conditions imposed on the observational data, we selected particles that satisfy $\lambda_c > 0.9$ and $| z |<0.3\kpc$. For this purpose, we employed {\sc agama} \citep{Vasiliev2019} to compute the necessary gravitational potentials for determining $\lambda_c$. These selection criteria ensure that our sample consists of kinematically cool particles in the geometrically thin disc, effectively excluding particles that were accreted during the merger. To determine the stellar age profiles of the models, as well as the uncertainties associated with $R_{\rm min}$, we employed the same method as described in Section~\ref{sec:profile_mw} and Appendix \ref{sec:age_profile_modelling}, which we used for the observational data. 

In Fig.~\ref{fig:age_fits_all_models}, we present the ML fit for each model (cyan curve). We also show the mean (red circles) and median (green triangles) age profiles for the models. The parameters describing the age profiles are summarised in Table~\ref{table:MW_age_prof_params}. The stellar age profiles obtained from the models show distinct shapes, reflecting the varied evolutionary paths of the models. The values for $R_{\rm min}$ are in the $ 9.54 <R_{\rm min}\,/\kpc < 14.04$ range and are thus comparable with the values obtained by our observational samples for the MW.

\begin{figure}
\includegraphics[width=\linewidth]{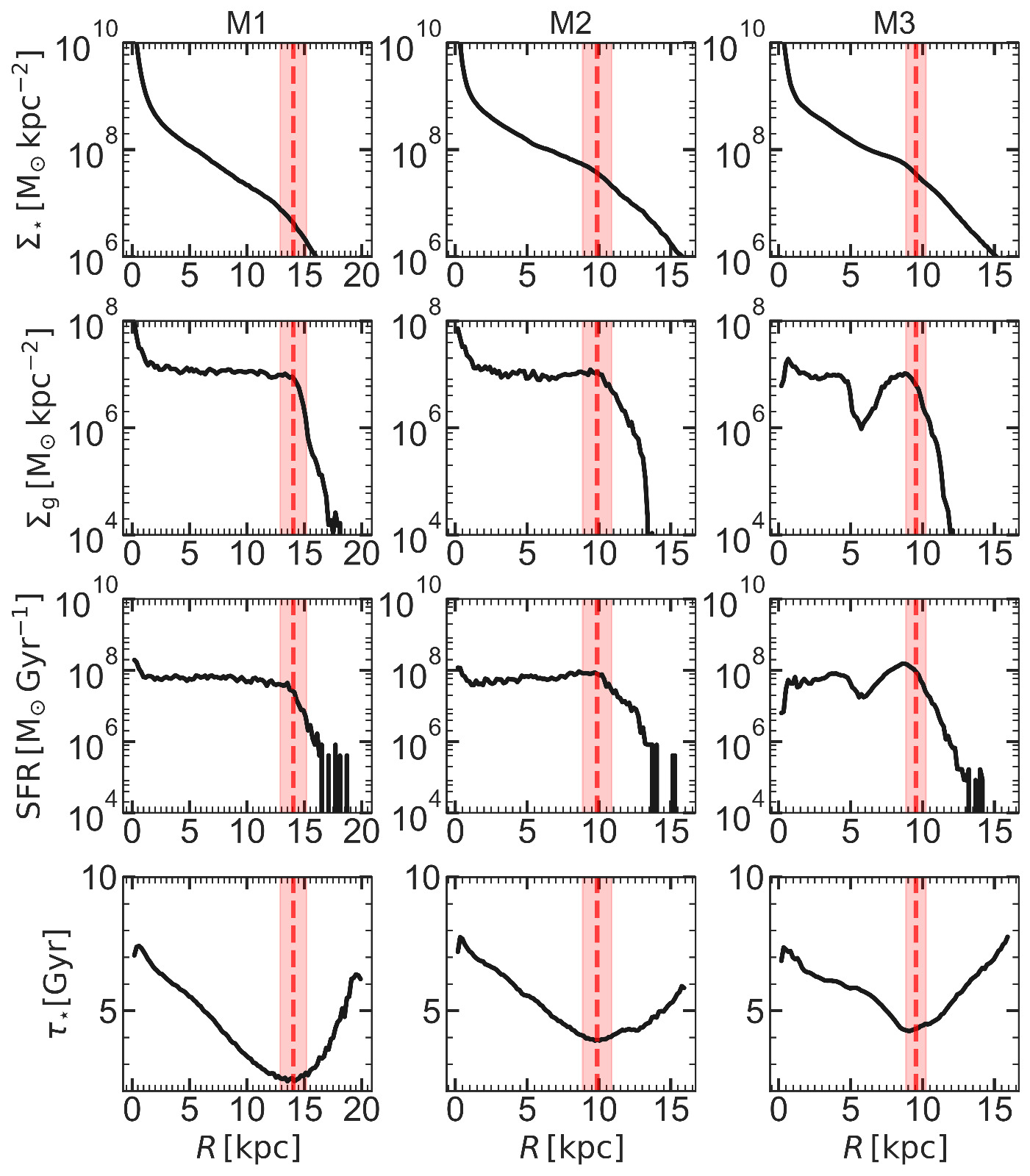}
    \caption{Top row: Radial stellar density profiles for the models. Second row: Cold gas $(T < 15,000\, \rm{K})$ density profile. Third row: Star formation rate profile. Bottom row: Mean stellar age profile. The vertical dashed red lines show the location of the age minimum for each model, with the red shaded region highlighting the uncertainty of the minimum. In each of our models, $R_{\rm min}$ coincides with a drop in the gas volume density, SFR, and the break in the stellar density profile (within the error).}
    \label{fig:model_breaks}
\end{figure}

\begin{figure*}
        \includegraphics[width=\linewidth]{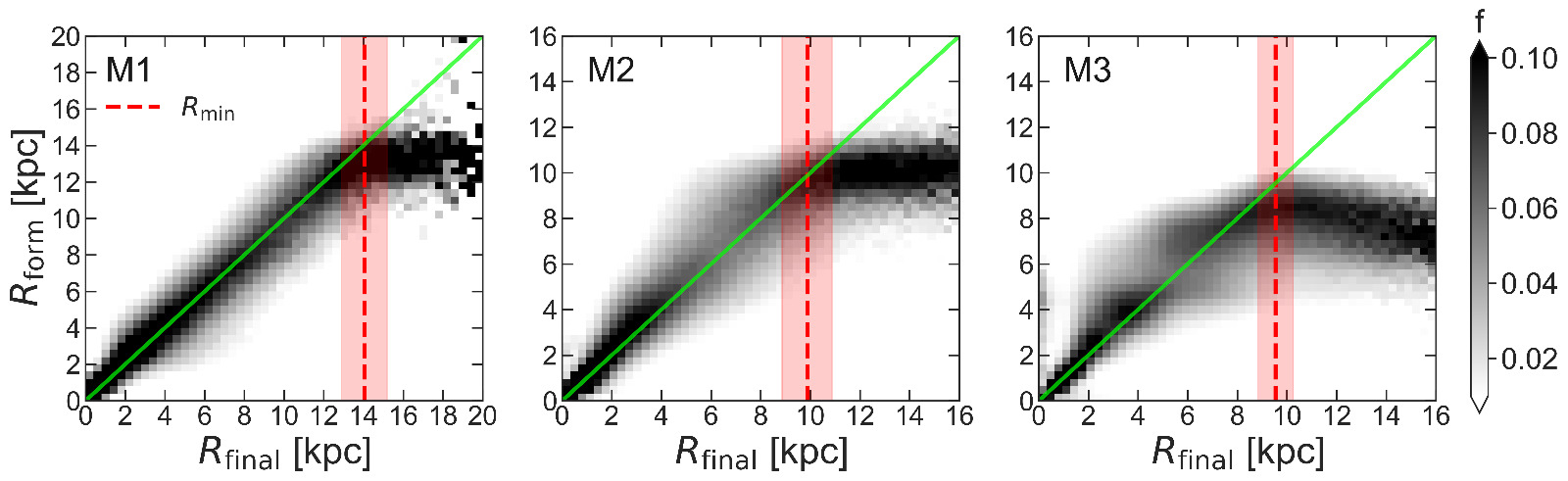}
    \centering
    \caption{Column-normalised 2D histogram of stellar birth radius versus final radius at $10\Gyr$ for the models. Stars here satisfy $|z|<0.3\kpc$ and $\lambda_{\rm c}>0.9$. The solid green and dashed red lines reflect the $1:1$ relation and the location of the age upturn, respectively. While the distribution follows the $1:1$ line in the inner disc, we find a knee in the distribution near the age upturn. Stars beyond the age upturn were born in the inner disc and subsequently migrated outwards.}
    \label{fig:migration}
\end{figure*}

In Fig.~\ref{fig:model_breaks}, we present the radial profiles of various quantities for the models at 10\Gyr. The first two columns show merger models, while the last column shows the isolated model. In the first row, we show the radial stellar density profiles for {\it \emph{all}} stellar particles, while the second row shows the density profile for cool gas $(T<15,000\, \mathrm{K})$. The third row shows the star formation rate in the disc. Lastly, the bottom row shows the stellar mean age profiles of the models. In all columns, the vertical dashed red line indicates the location of the age minimum, $R_{\rm min}$, determined by the minimum point of the ML fit. The red shaded regions reflect the uncertainty on $R_{\rm min}$. In all models, the age minimum roughly coincides with sharp truncations in the SFR profiles (within the uncertainty). In the inner disc ($R < R_{\rm min}$), these models have sufficient cold gas for star-formation, leading to a negative age gradient due to the inside-out growth. Conversely, for $R > R_{\rm min}$, the truncation in the cold-gas density profile leads to a sharp drop in the SFR. Despite this, the overall stellar profile (top row) is not sharply truncated at $R_{\rm min}$, but rather displays a Type~II (down-bending) break, leading to an outer disc region with a shorter scale length.

In Fig.~\ref{fig:migration}, we show column-normalised 2D histograms of the stellar birth radius, $R_{\rm form}$, versus final radius, $R_{\rm final}$, at $10 \Gyr$ for the same models shown in Fig.~\ref{fig:model_breaks}. This analysis is again restricted to stellar particles satisfying $\lambda_c > 0.9$ and $|z|<0.3\kpc$. The solid green line represents $R_{\rm form} = R_{\rm final}$, while the dashed red line reflects the location of the age minimum, $R_{\rm min}$. For stars with $R_{\rm final}<R_{\rm min}$, the distribution roughly follows the $1:1$ line, exhibiting some scatter indicative of random migration. In contrast, for $R_{\rm final}>R_{\rm min}$, the distribution diverges significantly from the $1:1$ line, with stellar particles in this region having a birth radius that is less than $R_{\rm min}$. In agreement with \citet{Roskar+2008a}, we confirm that stellar particles in the outer-disc region with $R >R_{\rm min}$ have been driven there via radial migration, leading to a positive age gradient in the outer disc.

\section{Discussion}\label{sec:discussion}

In the above analysis, we used the sub-sample of giants from LAMOST-DR3 from the \citet{Sanders+2018} value-added catalogue as well as giants from APOGEE-DR17 \citep{Abdurro+2022} with age estimations from {\tt astroNN} \citep{leung+2019} to reveal the edge of the MW star-forming disc. This was carried out by demonstrating that there is a positive age gradient in the anti-centre direction. We chose these samples for the reliability of the age estimations and for good coverage of the outer disc.

\subsection{Azimuthal variations in the stellar age profile}\label{sec:azimuthal_var}

\begin{figure*}
\includegraphics[width=\linewidth]{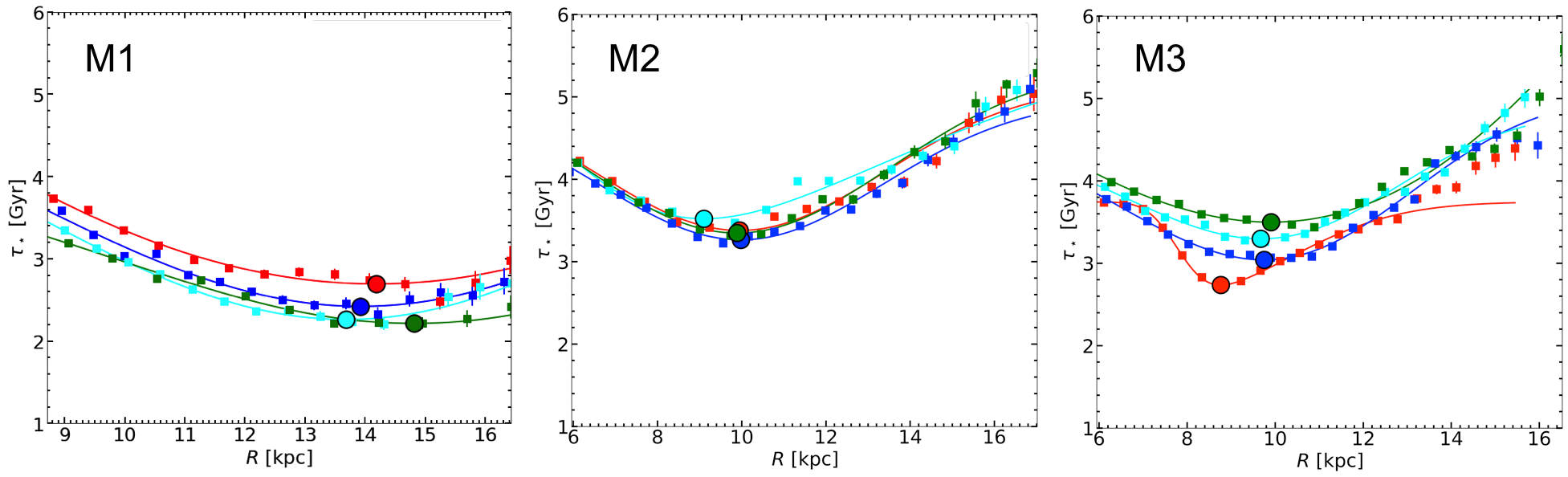}
    \centering
    \caption{Age profile fits for each quadrant for models M1, M2, and M3 (left to right). For each fit, the corresponding $R_{\rm min}$ is shown by the marker of the same colour. }
    \label{fig:quadrants}
\end{figure*}

The outskirts of disc galaxies, including the MW, commonly host a spiral structure, which can imprint spatial variations in the stellar populations, thereby affecting age and density profiles. In our simulation analysis, we exploited the full disc to determine the age profiles; by contrast, our MW profiles rely on observational datasets primarily in the anti-centre direction (see Fig.~\ref{fig:datasets_xy}). To assess azimuthal variations in the stellar age profiles, we partitioned each model into four quadrants (Q1–Q4) and compute the age profiles for each quadrant using the same fitting procedure employed throughout this work. Fig.~\ref{fig:quadrants} shows the quadrant-specific age profiles, with each curve corresponding to a distinct quadrant. Across all models, the azimuthal scatter in the derived $R_{\rm min}$ values is approximately $1\,\kpc$. This variation is smaller than the uncertainties reported in Section~\ref{sec:profile_mw} for the observational datasets ($\pm 0.75\,\kpc$ for LAMOST-DR3 and $\pm 0.82\,\kpc$ for APOGEE-DR17). We therefore conclude that azimuthal variations are unlikely to substantially bias the MW $R_{\rm min}$ values reported in this work.

\begin{figure}
\includegraphics[width=\linewidth]{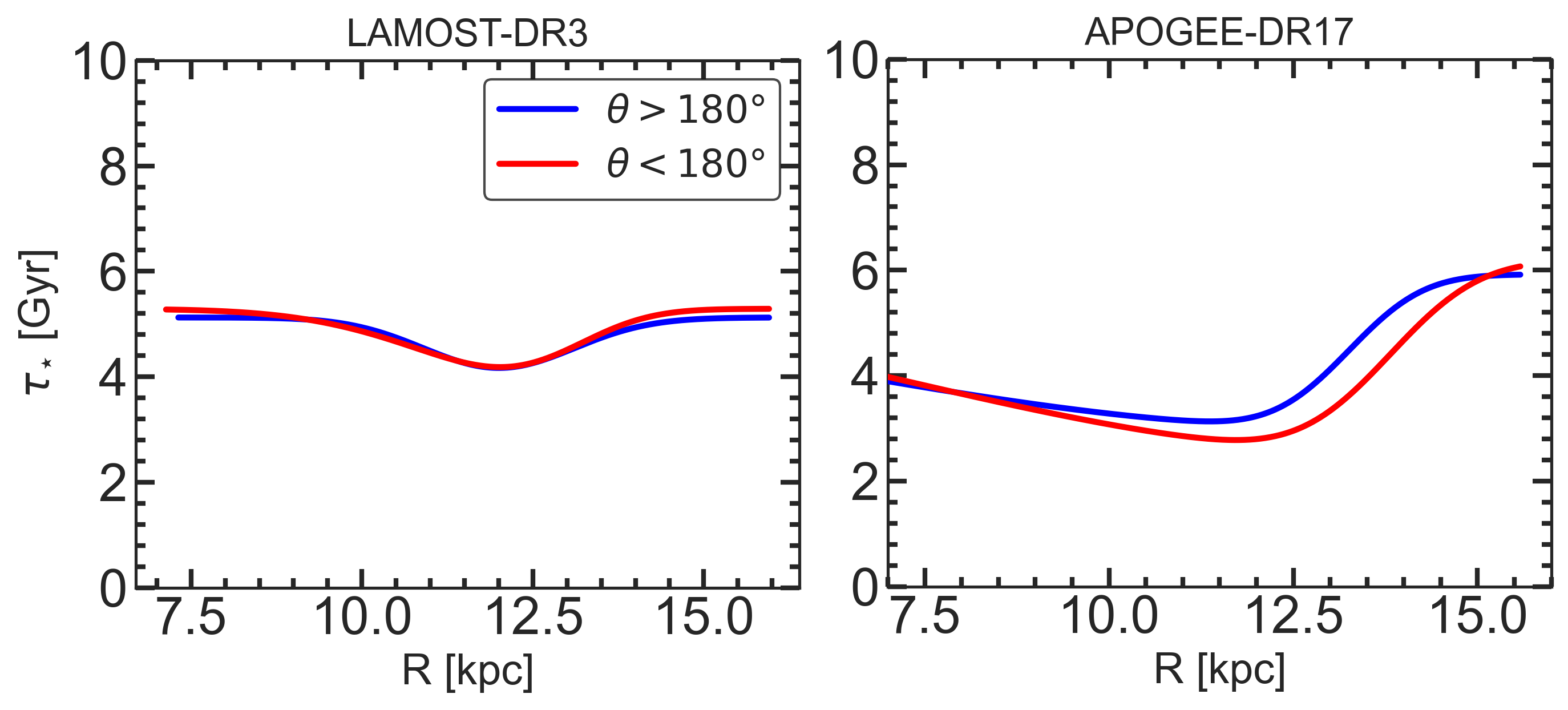}
    \centering
    \caption{Azimuthal age profile fits for the LAMOST-DR3 (left) and APOGEE-DR17 (right) datasets. We split each dataset into two azimuthally separated spatial bins to investigate spatial variations in the age profiles. For LAMOST-DR3 and APOGEE-DR17, the shift in $R_{\mathrm min}$ is approximately $100\pc$ and $400\pc,$ respectively. The higher shift in $R_{\mathrm min}$ for APOGEE likely reflects the smaller sample size, which makes the fitting process more difficult. Nonetheless, this is well within the uncertainty we report for the main samples in Sect.~\ref{sec:model_profiles}.}
    \label{fig:dataset_quadrants}
\end{figure}

To verify this conclusion observationally, we applied the same approach to our MW datasets by dividing them into two azimuthally separated spatial bins. Fig.~\ref{fig:dataset_quadrants} shows the resulting ML fits for LAMOST-DR3 (left) and APOGEE-DR17 (right), where the red and blue curves represent stars with $\theta > 180\degrees$ and $\theta < 180\degrees$, respectively. The observed age profiles show minimal variation between the two bins, with $R_{\rm min}$ differing by only $\sim120\,\pc$ for LAMOST-DR3 and $\sim400\,\pc$ for APOGEE-DR17. These small azimuthal variations are well within the uncertainties of our fitted $R_{\rm min}$ values in Sect.~\ref{sec:profile_mw} and further support our conclusion that azimuthal structure does not significantly affect the determination of the age minimum radius in the MW disc.

\subsection{Linking U-shaped age profiles and disc breaks}\label{sec:agemin_typeII}

The presence of a positive mean age gradient in the MW's outer stellar disc has important implications for our understanding of the large-scale structure of the disc and the processes underlying the formation of its outer regions. In Sect.~\ref{sec:model_profiles}, we describe how we used N-body+SPH simulations of both isolated and merger simulations to demonstrate that the observed positive age gradient in the kinematically cool outer disc can arise from a combination of a rapid star-formation drop and radial migration. In agreement with earlier numerical studies \citep[e.g.][]{Roskar+2008a}, our results indicate that this mechanism leads to the formation of a Type II (down-bending) stellar disc. The transition from inner to outer disc is marked by a break in the stellar density profile and reflects the edge of the star-forming disc. Crucially, we demonstrate that for the models, the minimum in the age profile ($R_{\rm min}$) coincides (within uncertainties) with the break in the stellar density profile for all stars.

The link between U-shaped age profiles and Type II discs is supported by observations. In a study of 85 late type spirals, \citet{Bakos+2008} found that 39 galaxies in their sample had a Type II light profile. All the Type II galaxies in their sample showed U-shaped colour profiles, with the minima in colour being coincident with the breaks in the light profiles. \citet{Chamba+2022} identified breaks in 273 disc galaxies in SDSS data and similarly found that the minimum in the colour profile generally coincided with the break in the disc's light profile. At higher redshifts, \citet{Azzollini+2008} observed similar trends in a sample of 232 Type II disc galaxies at redshifts of $0.1<z<1.1$. In more recent work, based on 233 disc galaxies at redshifts $z = 1 - 3$ observed by the {\it JWST}, \citet{Yu+2024} found that the minima in the colour profiles are coincident with the break in the Type II discs.

\subsection{Comparison with other MW studies}

As demonstrated in Sect.~\ref{sec:model_profiles}, our models indicate that the minimum in the age profile coincides with the edge of the star-forming disc. When combined with outward stellar migration, this produces a downward break in the stellar density profile. The age upturn observed in the LAMOST-DR3 and APOGEE-DR17 datasets therefore suggests a break in the MW disc stellar density at approximately $R_{\rm br} = 10.65-12.2\kpc$.

Our estimated break radius is smaller when compared to early studies, which generally report breaks in the $12.4 < R\, [\kpc] < 15$ range \citep[e.g.][]{Robin+1992, Ruphy+1996, Freudenreich+1998, Sale+2010, Minniti+2011}. The discrepancy may arise due to several factors, including differences in the assumed solar position and a limited number of sources in earlier catalogues, particularly in the anti-centre direction. 

By contrast, our results are more consistent with recent observational studies using mono-abundance populations (MAPS) from APOGEE \citep{Mackereth+2017,Bovy+2016} and LAMOST \citep{Yu+2021}. In particular, \citet{Mackereth+2017} and \citet{Bovy+2016} both report an anti-correlation between $R_{\rm br}$ and $\feh$ for low-$\alpha$ stars \citep[but see][]{Yu+2021}, identifying an outermost break in the $R=10 - 13\kpc$ region. Independently, \citet{Lian+2022b} used MAPs from APOGEE-DR17 to derive the spatial intrinsic density distribution of the MW disc. They explicitly find a broken radial density profile for both high- and low-$\alpha$ MAPs, with the break radius at $\approx 11\kpc$ in the low-$\alpha$ MAPs. Moreover, subsequent studies using APOGEE-DR17 and \gaia-DR3 data, \citet{Haywood+2024}, \citet{Lian+2024}, and \citet{Imig+2025} all independently report an outer break radius of $R_{\rm br} \sim 12\kpc$ in low-$\alpha$ populations.

Our reported location of the age upturn is consistent with previous studies that noted a positive age gradient in the outer MW disc. Specifically, \citet{Johnson+2025a} reported evidence of a slight increase in stellar ages beyond $R > 10~\mathrm{kpc}$ (see their Figure 3), based on the analysis of red giant stars from APOGEE-DR17 and \textsc{astronn} stellar age estimates \citep[see also][]{Johnson+2025b}.

The negative age gradient inside the break radius of the MW suggests inside-out growth of the disc within this radius. This interpretation is consistent with \citet{Katz+2021}, which used ages for a sample of $199, 307$ giant stars from APOGEE-DR16 and found evidence of inside-out growth up to a galactocentric radius of $R=10-12\kpc$ (see their Figure 11). Independently, \citet{Elia+2022}, using dense stellar clumps identified in the Hi-GAL Galactic plane survey, derived the MW’s SFR profile as a function of galactocentric radius up to $R=16\kpc$. As shown by their cumulative SFR profile (see their Figure 6), the profile reaches $\sim100\%$ of the total SFR at $11$–$12\kpc$, indicating that the majority of star formation occurs within this radius.

\subsection{Determining factor of the break radius in the Galaxy}

A key question that arises concerns what determines the break radius, $R_{\rm br}$, in the Galactic disc. One possibility is that the break radius is linked with the bar, which can have a strong influence on the structure of the host galaxy by inducing radial redistribution of gas \citep[e.g.][]{Kormendy+2013}. \citet{Munoz-mateos+2013} investigated the structural characteristics of 218 nearby, face-on galaxies using deep $3.6 \mu m$ imaging data from the \textit{Spitzer} Survey of Stellar Structure in Galaxies (S$^4$G). The authors sought to explore the connection between $R_{\rm br}$ in Type II discs and various structural parameters. In particular, they examined whether the bar radius, $R_{\rm bar}$, influences $R_{\rm br}$. Their analysis revealed that for galaxies with stellar masses exceeding $10^{10}M_{\odot}$, the distribution of the ratio $R_{\rm br}/R_{\rm bar}$ is bimodal (see their Figure 12), exhibiting two distinct sequences centered around $R_{\rm br}/R_{\rm bar} \sim 2-3$ and $\sim 3.5$. They argued that the sequence at $R_{\rm br}/R_{\rm bar} \sim 2-3$, also identified in previous studies \citep[e.g.][]{Erwin+2008,Pohlen+2006}, is linked with the outer Lindblad resonance (OLR) of the bar. We can apply this analysis to the Galaxy by assuming a bar radius of $R_{\rm bar} = 5\kpc$ \citep{Bland+2016}, which gives $R_{\rm br}/R_{\rm bar} = 2.3$ and $2.4$ for the APOGEE-DR17 and LAMOST-DR3 samples, respectively. This places the MW within the sequence where the break may be linked to the bar's OLR. This is also consistent with \citet{Clarke+2022}, which derived a bar OLR radius of $10.7 < R_{\rm OLR} \rm{(kpc)} < 12.4$. \citet{Haywood+2024} also made the connection between the reported outer break at $R=10-12\kpc$ and the bar's OLR, interpreting a sharp drop in the metallicity profile of stars around this radius as the dynamical effect of the OLR.

Alternatively, star-formation thresholds regulated
by thermal instabilities arising from the transition between warm
and cold gas phases may also play a role \citep[e.g.][]{Kennicutt+1989, Schaye+2004, Elmegren+2006, Martin+2012}. \citet{Schaye+2004} showed, using theoretical models, that when the local gas density falls below the threshold for the cold-phase transition, the gas remains warm and stable against large-scale collapse under the influence of the ambient UV background, thereby suppressing star formation. This naturally produces a break in the star-formation and stellar surface-density profiles at radii where the cold-phase threshold is crossed.

Lastly, the star-formation drop may be linked to the warp in the Galactic disc. In an observational study of 23 edge-on galaxies, \citet{Kruit+2007} found that $17$ galaxies in the sample showed evidence of a break, with $R_{\rm br}$ roughly coincident with the onset of the warp ($R_{\rm warp} \approx 1.1R_{\rm br}$). \citet{Sanchez+2009} used a cosmological hydrodynamical model of a disc galaxy to study the source of Type II discs seen in many galaxies. The authors found a U-shaped stellar age profile in their model, with a minimum that coincided with the break in the surface brightness profile. The authors found that the drop in gas volume density (and therefore SFR) in their model is due to a warp in the gaseous disc. Indeed, they also demonstrated that $(R_{\rm br})$ in their model coincides with the onset of the gas warp. Observational studies of external galaxies also find drops in the HI gas density, which coincides with the onset of warps \citep{Garcia+2002,Jozsa+2007}. 

There is an extensive and well-established body of observational evidence confirming the presence of a warp in the MW, observed consistently across multiple tracers -- including the HI \citep{Kerr+1957,Burke+1957,Weaver+1974,Levine+2006,Kalberla+2007} and in the stellar disc \citep{Djorgovski+1989,Porcel+1995,Freudenreich+1998,Drimmel+2001,Lopez+2002}. Recent large-scale surveys, including \gaia, have mapped the Galactic warp in greater detail \citep[e.g.][]{Chen+2019,Skowron+2019,Cheng+2020, Poggio+2020, Chrobakova+2022,Lemasle+2022, Dehnen+2023, Zhou+2024, Cabrera-gadea+2024,Cabrera-gadea+2024b,Jonsson+2024, Poggio+2025}. Studies have sought to constrain the geometry and kinematics of the warp using variable stars as tracers. Being a very young and therefore kinematically cool population, Classical Cepheid variables are ideal candidates for studying the warp, as it is expected that they trace the geometry of the HI gas from which they formed. \citet{Skowron+2019} and \citet{Lemasle+2022} both used Classical Cepheids from OGLE to argue a very short warp onset of $R \approx 4-5\kpc$. In contrast, \citet{Chen+2019} derived a warp onset of $R \approx 10\kpc$ using $1300$ Cepheids from WISE. Similarly, \citet{Dehnen+2023} used a sample of Cepheid variables with full 6D coordinates from \gaia-DR3 and found that the warping commences at $R \approx 11\kpc$, with the stellar disc becoming increasingly inclined at larger radii. A warp onset in the outer disc is also consistent with studies using other types of variables. \citet{Cabrera-gadea+2024} examined the spatial properties of metal-rich RR Lyrae variables from a combined catalogue of \gaia\ and other photometric surveys, finding a kinematic signal of the warp at $R \approx 10\kpc$. If the onset of warps is correlated with disc break radii, as found by \citet{Kruit+2007}, then we would expect the stellar break in the MW to occur close to the $R=10-11\kpc$ region. The age minimum in the MW disc we observed in this study is therefore also consistent with a warp onset in the outer disc region ($R>10\kpc$). 

In the MW, the root cause of the star-formation drop beyond the break radius remains unclear. The break could be primarily driven by bar-related dynamics and the OLR, by a thermally regulated transition in gas phases, by the onset of a warp in the gaseous and stellar discs, or by a combination of these mechanisms operating together. The combination of detailed numerical studies and upcoming data from surveys such as \gaia-DR4 will allow us to obtain a clearer picture.

\section{Summary}\label{sec:summary}

We employed spectroscopic data from APOGEE-DR17+{\tt astroNN} and LAMOST-DR3 to explore the stellar age distribution in the MW disc. In order to disentangle churning from blurring, we required stars be close to the midplane ($|z|<0.3\kpc$) and on roughly circular orbits ($\lambda_{\rm c}>0.9$). We list our conclusions below.

\begin{itemize}

  \item Our findings indicate that the outer MW disc is predominantly kinematically cold. Applying a circularity cut ($\lambda_{\rm c}>0.9$) to disc stars from the LAMOST-DR3 and APOGEE-DR17 ($|z|<0.3\kpc$) sources removes only a small fraction of stars from our samples. Indeed, over $90\%$ of disc stars in the outer disc ($R > 11.5\kpc$) are on cold orbits (see Sect.~\ref{sec:sample_properties}). Crucially, our selection criteria successfully exclude halo stars, as demonstrated through Toomre diagrams (Fig.~\ref{fig:lamost_chr}) and orbital anisotropy measurements ($\beta < 0$), confirming that our sample consists of tangentially biased disc stars rather than radially biased halo populations.

  \item We find that the Galaxy has a U-shaped stellar age profile, where a negative stellar age gradient leads to a positive gradient in the outer disc. Both APOGEE-DR17 and LAMOST-DR3 samples show consistent age profiles across the disc: a negative age gradient from $7.0\mhyphen10.0\kpc$, a relatively flat plateau-like region from $10\mhyphen12\kpc$, and a clear positive age gradient beyond $12\kpc$ (Fig.~\ref{fig:obs_fit_ML}). We identify the location of the age minimum at $R_{\rm min}=11.28\pm0.58\kpc$ and $R_{\rm min}=12.15\pm0.62 \kpc$ for the APOGEE-DR17 and LAMOST-DR3 samples, respectively.

  \item Using N-body+SPH simulations, we demonstrate that $R_{\rm min}$ corresponds to the edge of the Galaxy's star-forming disc, where there is a sharp drop in the SFR (see Sec.~\ref{sec:model_profiles}). While we are unable to distinguish between potential physical drivers of this SFR drop --such as the bar's outer Lindblad resonance (OLR), the Galactic warp, or extragalactic ionising fields from the intergalactic medium-- we conclude that the combination of a drop in SFR at $R_{\rm min}$ and outward stellar migration beyond this radius drives the observed U-shaped age profile. We argue that the presence of such a profile indicates that the Galaxy has a Type II (down-bending) stellar density profile, where the break in the density profile coincides with $R_{\rm min}$. Our reported values for  $R_{\rm min}$ are consistent with break radii reported for low-$\alpha$ populations in the MW disc.

\end{itemize}
 
\begin{acknowledgements}
We thank the anonymous referee for providing feedback which improved this work.
KF acknowledges financial support from the European Research Council under the ERC Starting Grant “GalFlow” (grant 101116226).
JC acknowledges support by Xjenza Malta (formerly, the Malta Council for Science and Technology, MCST) through IPAS 2019-028.
SGK was supported by the Moses Holden Studentship.
JA is supported by the National Natural Science Foundation of China under grant Nos. 12233001, 12533004, by the National Key R\&D Program of China under grant No. 2024YFA1611602, by a Shanghai Natural Science Research Grant (24ZR1491200), by the ``111'' project of the Ministry of Education under grant No. B20019, and by the China Manned Space Program with grant Nos. CMS-CSST-2025-A08, CMS-CSST-2025-A09 and CMS-CSST-2025-A11. 
LBeS acknowledges support from CNPq (Brazil) through a research productivity fellowship, grant no. [304873/2025-0].
TK acknowledges support from the NSFC (Grant No. 12303013) and support from the China Postdoctoral Science Foundation (Grant No. 2023M732250)
VC acknowledges support from the Agencia Nacional de Investigaci\'{o}n y Desarrollo (Chile) through the FONDECYT Iniciaci\'{o}n Grant (No. 11250723).
This work has made use of data from the European Space Agency (ESA) mission \gaia\ (\url{https://www.cosmos.esa.int/gaia}), processed by the \gaia\ Data Processing and Analysis Consortium (DPAC, \url{https://www.cosmos.esa.int/web/gaia/dpac/consortium}). 
Funding for the DPAC has been provided by national institutions, in particular the institutions participating in the \gaia\ Multilateral Agreement. 
The simulations in this paper were run both at the High Performance Computing Facility of the University of Lancashire and at the DiRAC Shared Memory Processing system at the University of Cambridge,
operated by the COSMOS Project at the Department of
Applied Mathematics and Theoretical Physics on behalf of the STFC DiRAC High Performance Computing (HPC) Facility: www.dirac.ac.uk. This equipment was funded by the Department for Business, Innovation and Skills (BIS) National E-infrastructure capital grant ST/J005673/1, STFC capital grant ST/H008586/1, and STFC DiRAC Operations grant ST/K00333X/1. DiRAC is part of the National E-Infrastructure. 
\end{acknowledgements}

\bibliographystyle{aa}
\bibliography{refs_mw} 

\begin{appendix}

\section{Modelling Age Profiles}
\label{sec:age_profile_modelling}

The overall age profile modelling method is the same for the MW and simulation datasets. For each dataset, we select stars with $\lambda_c>0.9$ and $|z|<0.3\kpc$ to ensure we analyse stars with (largely) circular orbits in the mid-plane. We construct the ($\tau_{\star}, R$)-plane with 30 bins in $R$, with $5<R/\mathrm{kpc}<R_\mathrm{outer}$, where $R_\mathrm{outer}$ is defined as the first radial bin where the number of stars $<50$. We ignore the inner 5 kpc to avoid age profile complexity due to the bar in the MW, which would complicate the fits unnecessarily. We plot the binned mean profile, then smooth the resulting profile with a second order Butterworth filter \citep{butterworth_1930} followed by cubic spline interpolation, and use the Python signal processing library\footnote{The {\fontfamily{qcr}\selectfont{find\_peaks}} routine.} to identify the radial location of the minimum in $\tau$, denoted $R_{\rm min}$. We then model the $\tau_{\star}-R$ relation for all radial bins.

After experimentation with various functional forms, we found that an inverted skewed Gaussian of the following form
\begin{equation}
    \tau(R)=-a\exp[-c(R-b)^2][1+\mathrm{erf}(d(R-b))] + \epsilon,
    \label{eqn:skewed-Gaussian}
\end{equation}
gives the best fits to the data. The function asymptotes to $\epsilon$ as $R\xrightarrow{}\infty$. The parameter $d$ controls the extent of the asymmetry ($d=0$ means symmetric, and the function reverts to an inverted Gaussian), and $b$ ensures that the minimum in $R$ is offset from $R=0$.

We model $\tau(R)_{\star}$ using a maximum-likelihood method. $\tau(R|\Theta)$ is a star's age as a function of Galactocentric radius $R$, given the fixed parameter set $\Theta = \{a, b, c, d, \epsilon\}$. The `likelihood function' ($\ie$ the probability of the data for a given $\Theta$) is defined as

\begin{equation}
    \mathcal{L}(\Theta)=\prod_{i=1}^{N}p_i(\tau_i,{R_i}, \sigma_i
    |\Theta),
    \label{eqn:likelihood}
\end{equation}
the product running over all $N$ stars in the radial cut. $p_i$ is the probability that a particle $i$ has age $\tau_i$ and radius $R_i$ for a parameter set $\Theta$. $\sigma_i$ is the uncertainty in the age.

For the observations, age uncertainties are provided. For the models, this is not the case. We wish to model the uncertainty in the age of each particle $\tau_i$, since this is the dominant source of uncertainty in the observations. To this end, in the models, $\sigma_i$ is drawn from a normal distribution with mean equal to the overall dispersion in age amongst the stars (this is $\sim 2$ Gyr in the models), and dispersion $0.1$ Gyr. The uncertainty in $R$ for each particle is assumed to be negligible as is the case for the observations, and we do not model it.

To find the optimal parameter set, we maximise the likelihood. It is more mathematically convenient to take the logarithm of Equation~\ref{eqn:likelihood}, as the logarithm is a monotonically increasing function of its operand, maximising it will maximise $\mathcal{L}(\Theta)$. The logarithm yields
\begin{equation}
    \ln{\mathcal{L}(\Theta)}=\sum_{i=1}^{N}\ln{[p_i(\tau_i,{R_i}, \sigma_i|\Theta)]}.
    \label{eqn:log-likelihood}
\end{equation}
Assuming a Gaussian distribution of the data as follows: 

\begin{equation}
    p_i(\tau_i,{R_i}, \sigma_i)=\frac{1}{\sigma_i \sqrt{2 \pi}} \exp \left(-\frac{[\tau_i-\tau(R_i)]^2}{2 \sigma_i^2}\right),
    \label{eqn:gen_Gaussian}
\end{equation}
with $\tau(R_i)$ given by Equation~\ref{eqn:skewed-Gaussian}, we must then find the set of parameters $a,b,c,d,\epsilon$ which maximise

\begin{equation}
    \ln{\mathcal{L}(\Theta)} = \sum_{i=1}^N -0.5[\ln(2\pi) - \ln(\sigma_i^2)] - [\tau_i - \tau(R_i)]^2/2\sigma_i^2].
    \label{eqn:to_maximize}
\end{equation}

The negative log likelihood, $\ln{\mathcal{L}(\Theta)}$, is minimised (thus maximising $\mathcal{L(\Theta)}$) using a Nelder-Mead downhill simplex algorithm (Python's \texttt{fmin} routine).

\begin{figure}
    \includegraphics[width=\linewidth]{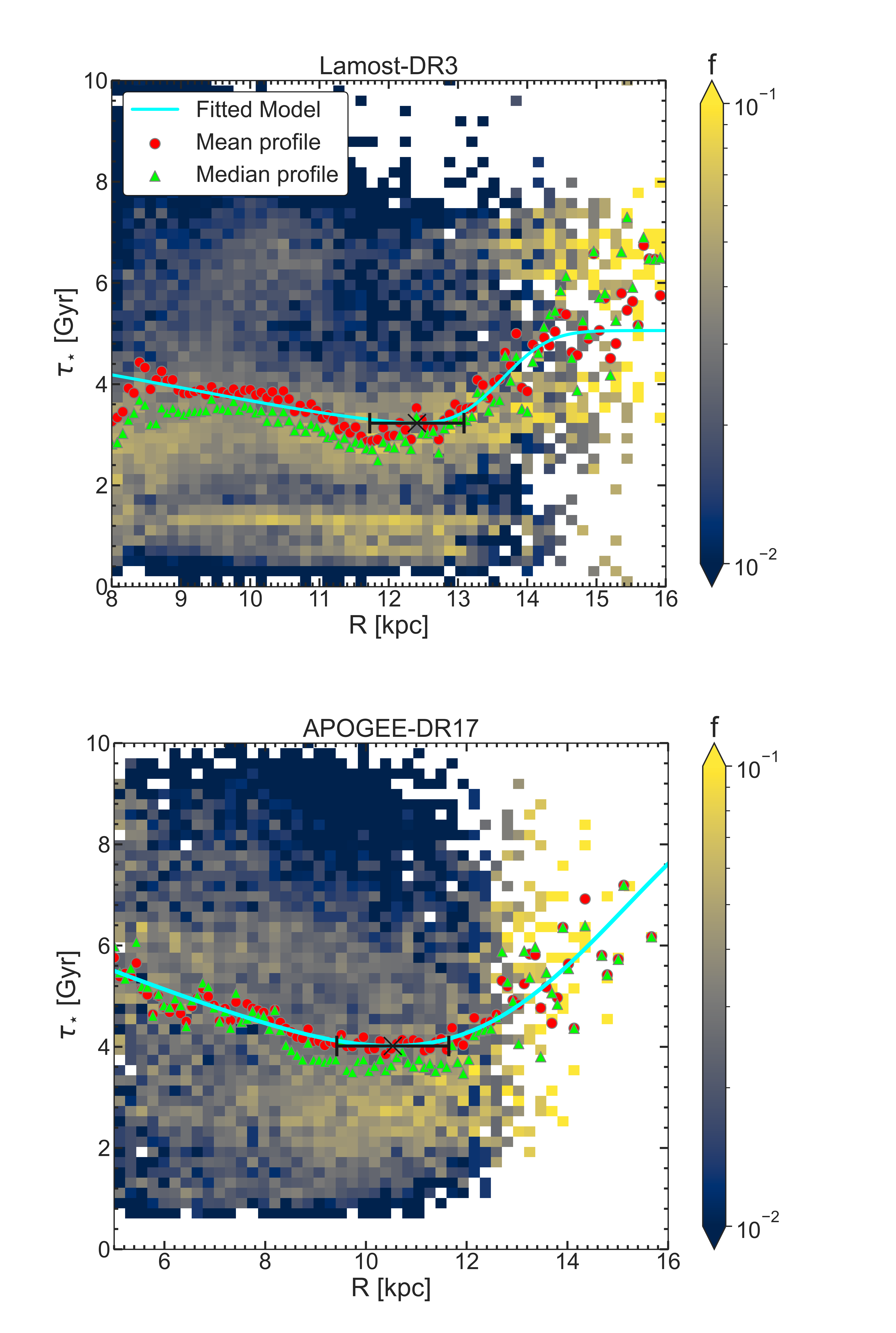}
    \caption{The stellar age profiles, $\tau_{\star}(R)$, for the APOGEE-DR17 (bottom) and LAMOST-DR3 (top) samples for stars with $\sigma_{\tau_{\star}}/\tau_{\star} < 0.3$.}
    \label{fig:mw_fit_30}
\end{figure}

\section{Models}

The models analysed in this work are all run with {\sc gasoline} \citep{Wadsleyetal2004, Wadsley2017} which is the SPH implementation of the N-body tree code {\tt PKDGRAV} \citep{stadel2001}. 
We use a base timestep of $\Delta t=5\Myr$ with timesteps refined such that $\delta t = \Delta t/2^n < \eta\sqrt{\epsilon/a_g}$. Here, we set the refinement parameter $\eta = 0.175$. The opening angle of the tree-code is set to $\theta_{a} = 0.7$. The gas particle timesteps also satisfy $\delta t_{\rm gas} = \eta_{\rm{courant}}h/[(1 + \alpha)c + \beta \mu_{\rm{max}}]$, where $\eta_{\rm{courant}} = 0.4$, $h$ is the SPH smoothing length set over the nearest 32 stars, $\alpha$ and $\beta$ are the linear and quadratic viscosity coefficients, and $\mu_{\rm{max}}$ is described in \citet{Wadsleyetal2004}. We describe the isolated model M3 in Section~\ref{sec:model_iso}, and the models with a GSE merger in Section~\ref{sec:model_gastro}.

\subsection{Isolated model}\label{sec:model_iso}

The isolated model M3 is identical to the model m2\_c\_nb used in \citet{Fiteni+2021}, and is a higher mass resolution version of the simulation described in \citet{Clarke+19}, \citet{beraldo+2020}, and \citet{amarante2020}. The simulation is evolved for $10 \Gyr$ with {\sc gasoline} \citep{Wadsleyetal2004, Wadsley2017}, starting off as a corona of hot gas embedded inside a dark matter halo with a virial radius of $r_{200} \simeq 200 \kpc$ and a virial mass of $M_{200} = 10^{12} \Msun$. Both components comprised $5 \times 10^6$ particles each, with softening parameters of 
$\epsilon = 50 \pc$ and $\epsilon = 100 \pc$ for the gas and dark matter respectively. At $t=0$, no stars are present. The gas corona has the same initial radial profile as the dark matter halo, but contains only $10\%$ of the mass. Gas particles are given a tangential velocity resulting in a spin parameter $\lambda \equiv J |E|^{1/2}/(GM_{\mathrm{vir}}^{5/2}) = 0.065$, where $J$ and $E$ are the total angular momentum and the energy of the gas particles, and $G$ is the gravitational constant \citep[e.g.][]{Peebles1969}.

Model M3 includes metal-line cooling \citep{ShenWadsleyStinson10}, allowing the gas to cool more efficiently. This, along with a low feedback model, results in an early episode of clump formation. The conditions for star-formation to occur require that the cold gas ($T < 15,000$ K) density exceeds 0.1 cm$^{-3}$ and is part of a converging flow. Star particles forming from the cooling gas also have a softening parameter of $\epsilon = 50 \pc$. The star-formation efficiency is set to $5\%$, and we use the feedback recipe described by \citet{stinson2006}. The resulting supernova feedback injects $10\%$ of the $10^{51}$ erg per supernova back into the interstellar medium. We also use turbulent diffusion for gas mixing as described by \citet{ShenWadsleyStinson10}.

\subsection{Models with a dwarf galaxy merger}\label{sec:model_gastro}

\citet{amarante+2022} introduced the Gaia-Enceladus-Sausage Timing, Chemistry and Orbit (GASTRO) library which contains a set of models where a GSE-like satellite merges into a MW-like galaxy during its first Gyrs of evolution. The merger models used in the present work are a subset of the GASTRO library. The input physics used in the models is fully described in \citet{amarante+2022}. The two models only differ in the total energy injected per supernova to the interestellar medium: $80\%$ and $20\%$ of the $10^{51}$ erg per supernova for model M1 and M2, respectively. Here, we describe the satellite's setup and its orbit, and summarize some key properties in Table \ref{table:gastro-models}. For the MW galaxy, the initial condition is the same as used for the isolated models described in Appendix \ref{sec:model_iso}. \par

Their initial conditions are generated with the \textsc{GalactICS} code \citep{kuijken-dubinski1995, widrow-dubinsky2005, widrow+2008} which allows generating equilibrium exponential gas discs \citep{deg+2019}. The setup is such that the dwarfs start with a Navarro-Frenk-White (NFW) DM profile and a gaseous disc, each containing $10^5$ and $2\times10^4$ particles, respectively. Their NFW halo has a scale velocity of $200\kms$ and is truncated at $50\kpc$. The gas disc in M1 (M2) has an exponential scale radii and initial gas mass of 1 kpc (5 kpc) and $1.4\times 10^9\, {\rm M_{\odot}}$ ($2.75\times 10^9\, {\rm M_{\odot}}$), respectively.

In both models the satellite's initial orbit circularity\footnote{Circularity of an orbit is defined in Section \ref{sec:datasets}} is $C=0.5$, their orbit inclination relative to the host galaxy's mid-plane is $15^{\circ}$, and they are on prograde orbits relative to the host' disc sense of rotation. They differ on their initial position: in M1 the satellite starts at a radial distance of 200 kpc from the host galaxy, whereas in M2 the initial distance is 150 kpc. Consequently, the merger in M1 and M2 is completed at $t \approx 3.6$ Gyr and $t \approx 2.6$ Gyr, respectively. Finally, the satellite's total stellar mass when the merger is completed is $1.4\times 10^8\, {\rm M_{\odot}}$ and $7.7\times 10^8\, {\rm M_{\odot}}$, for models M1 and M2, respectively.

\begin{table}
\caption{\label{table:gastro-models}Merger models properties}
\centering
\begin{tabular}{lcccc}
\hline\hline
Model & $t_{m}$ [Gyr] & ${\rm M_{g,0}}$ [${\rm M_{\odot}}$] & ${\rm M_{s,t_m}}$ [${\rm M_{\odot}}$] & FB (\%) \\
\hline
M1 & 3.6 & $1.4\times10^9$ & $1.4\times10^8$ & 80 \\
M2 & 2.6 & $2.75\times10^9$ & $7.7\times10^8$ & 20 \\
\hline
\end{tabular}
\tablefoot{The merger models presented in this work. The satellite in all the models are on prograde orbit with initial orbital circularity $\mathcal{C}=0.5$. We indicate the time when the merger is complete, $t_m$, the initial gas mass of the satellite, ${\rm M_{g,0}}$, and the total stellar mass at $t=t_m$, ${\rm M_{s,t_m}}$. Finally, FB is the amount of the $10^{51}$ erg supernova energy injected into the interstellar medium in each model.}
\end{table}

\end{appendix}

\end{document}